\documentclass[times]{aastex631}
\usepackage{gensymb}
\usepackage{placeins}
\usepackage{booktabs}
\usepackage{float}
\usepackage{color}
\usepackage[yyyymmdd,hhmmss]{datetime}
\newcommand{\CaIR}{Ca~{\sc{ii}}~8542{\,}{\AA}~}
\newcommand{\vlos}{$v_{\mathrm{LOS}}$}
\defcitealias{GM21}{\scshape GM21}

\shorttitle{Coherent Pore Waves}
\shortauthors{Grant et~al.}

\begin{document}

\title{The Propagation of Coherent Waves Across Multiple Solar Magnetic Pores}

\correspondingauthor{S.~D.~T. Grant}
\email{samuel.grant@qub.ac.uk}

\author[0000-0001-5170-9747]{S.~D.~T. Grant}
\affiliation{Astrophysics Research Centre, School of Mathematics and Physics, Queen’s University Belfast, Belfast, BT7 1NN, UK}

\author[0000-0002-9155-8039]{D.~B. Jess}
\affiliation{Astrophysics Research Centre, School of Mathematics and Physics, Queen’s University Belfast, Belfast, BT7 1NN, UK}
\affiliation{Department of Physics and Astronomy, California State University Northridge, Northridge, CA 91330, USA}

\author[0000-0002-5365-7546]{M. Stangalini}
\affiliation{ASI, Italian Space Agency, Via del Politecnico snc, 00133, Rome, Italy}

\author[0000-0002-7711-5397]{S. Jafarzadeh}
\affiliation{Leibniz Institute for Solar Physics (KIS), Sch{\"{o}}neckstr. 6, 79104 Freiburg, Germany}
\affiliation{Rosseland Centre for Solar Physics, University of Oslo, P.O. Box 1029 Blindern, 0315 Oslo, Norway}

\author[0000-0002-0893-7346]{V. Fedun}
\affiliation{Plasma Dynamics Group, Department of Automatic Control and Systems Engineering, The University of Sheffield, Mappin Street, Sheffield, S1 3JD, UK}

\author[0000-0002-9546-2368]{G. Verth}
\affiliation{Plasma Dynamics Group, School of Mathematics and Statistics, The University of Sheffield, Hicks Building, Hounsfield Road, Sheffield, S3 7RH, UK}

\author[0000-0001-8556-470X]{P.~H. Keys}
\affiliation{Astrophysics Research Centre, School of Mathematics and Physics, Queen’s University Belfast, Belfast, BT7 1NN, UK}

\author[0000-0003-0003-4561]{S.~P. Rajaguru}
\affiliation{Indian Institute of Astrophysics, Bangalore-34, India}

\author[0000-0002-2554-1351]{H. Uitenbroek}
\affiliation{National Solar Observatory, University of Colorado Boulder, 3665 Discovery Drive, Boulder, CO, 80303, USA}

\author[0000-0002-9901-8723]{C.~D. MacBride}
\affiliation{Astrophysics Research Centre, School of Mathematics and Physics, Queen’s University Belfast, Belfast, BT7 1NN, UK}

\author[0000-0001-9629-5250]{W. Bate}
\affiliation{Astrophysics Research Centre, School of Mathematics and Physics, Queen’s University Belfast, Belfast, BT7 1NN, UK}

\author[0000-0002-2593-4884]{C.~A. Gilchrist-Millar}
\affiliation{Astrophysics Research Centre, School of Mathematics and Physics, Queen’s University Belfast, Belfast, BT7 1NN, UK}

\begin{abstract}
Solar pores are efficient magnetic conduits for propagating magnetohydrodynamic wave energy into the outer regions of the solar atmosphere. Pore observations often contain isolated and/or unconnected structures, preventing the statistical examination of wave activity as a function of atmospheric height. Here, using high resolution observations acquired by the Dunn Solar Telescope, we examine photospheric and chromospheric wave signatures from a unique collection of magnetic pores originating from the same decaying sunspot. Wavelet analysis of high cadence photospheric imaging reveals the ubiquitous presence of slow sausage mode oscillations, coherent across all photospheric pores through comparisons of intensity and area fluctuations, producing statistically significant in-phase relationships. The universal nature of these waves allowed an investigation of whether the wave activity remained coherent as they propagate. Utilizing bi-sector Doppler velocity analysis of the \CaIR line, alongside comparisons of the modeled spectral response function, we find fine-scale 5 mHz power amplification as the waves propagate into the chromosphere. Phase angles approaching zero degrees between co-spatial line depths spanning different line depths indicate standing sausage modes following reflection against the transition region boundary. Fourier analysis of chromospheric velocities between neighboring pores reveals the annihilation of the wave coherency observed in the photosphere, with examination of the intensity and velocity signals from individual pores indicating they behave as fractured wave guides, rather than monolithic structures. Importantly, this work highlights that wave morphology with atmospheric height is highly complex, with vast differences observed at chromospheric layers, despite equivalent wave modes being introduced into similar pores in the photosphere.
\end{abstract}


\section{Introduction}

%%%%%%%%%%%%%%%%%%%%%%%%%%%%%%%%%%%%%%%%%%%%%%%
% FIGURE 1 
%%%%%%%%%%%%%%%%%%%%%%%%%%%%%%%%%%%%%%%%%%%%%%%
\begin{figure*}[t!]
  \centering
  \includegraphics[trim=15mm 0mm 30mm 0mm, clip, width=0.8\textwidth, angle=0]{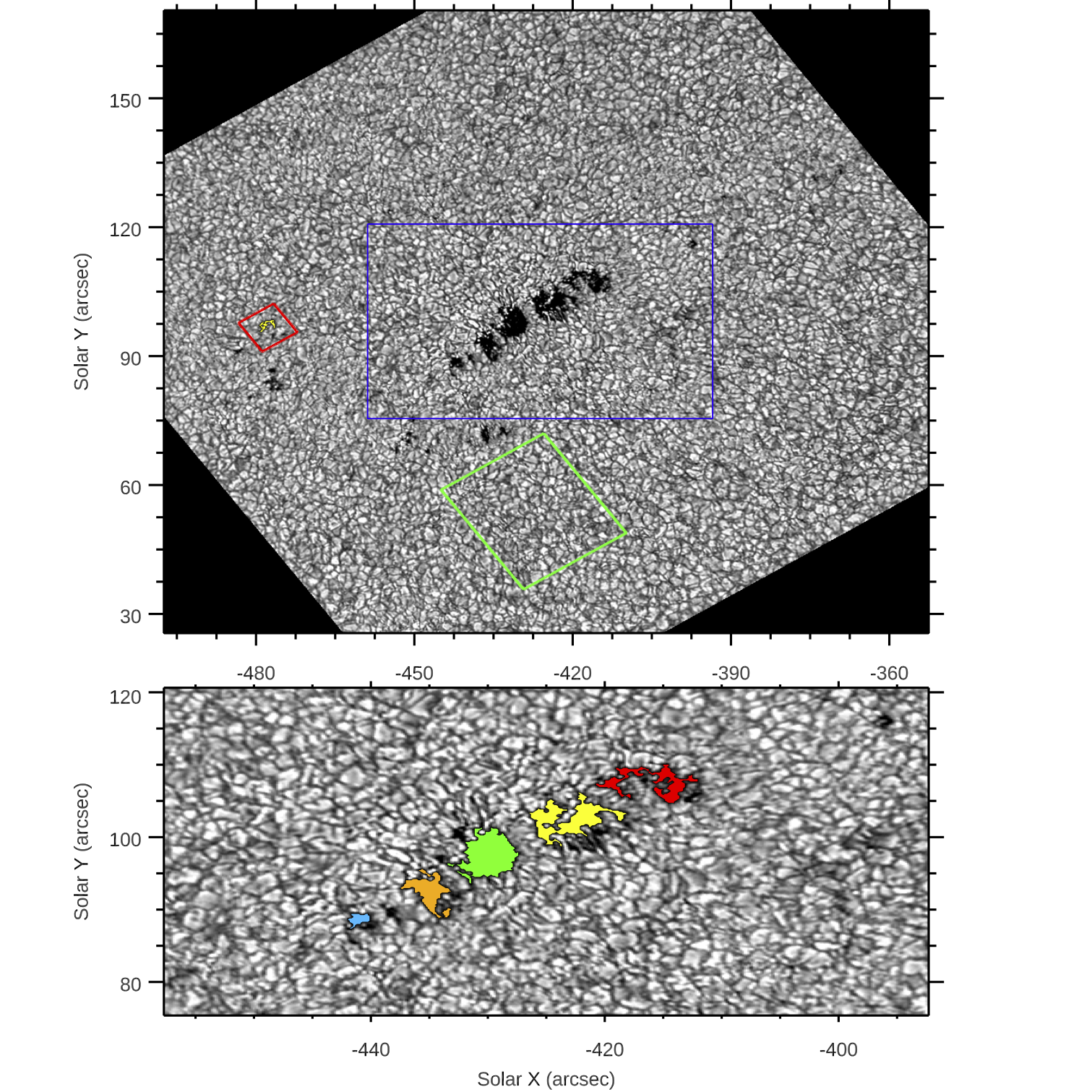}
\caption{The upper panel displays a $4170${\,}{\AA} continuum image of the full ROSA field-of-view in proper heliocentric coordinates. The green box outlines the quiescent region used for the calculation of intensity thresholds that are used to define the pore boundaries, the blue box highlights the region of interest displayed in the lower panel, and the red box contains the reference magnetic structure (yellow). The lower panel shows the five pores, labelled in ascending order from left-to-right, with pore 1 (P1) being blue, pore 2 (P2) orange, pore 3 (P3) green, pore 4 (P4) yellow, and pore 5 (P5) red.}
\label{ROSA_fov}
\end{figure*}
%%%%%%%%%%%%%%%%%%%%%%%%%%%%%%%%%%%%%%%%%%%%%%%
%%%%%%%%%%%%%%%%%%%%%%%%%%%%%%%%%%%%%%%%%%%%%%%
%%%%%%%%%%%%%%%%%%%%%%%%%%%%%%%%%%%%%%%%%%%%%%%

The mechanisms by which the upper solar atmosphere maintains its heightened temperature remain at the forefront of solar physics \citep{Parnell2012, VD2020}. The corona has long been the focus of this interest, due in part to the extraordinary temperatures observed, in excess of $1$~MK, but also to the availability of  a range of data products alongside models capable of replicating the physics of this optically thin region \citep{Ban2007, Moortel2015}. In contrast, the chromosphere is a challenging observational project, with less than $1\%$ of observed solar emission emanating from this tenuous region \citep{Jess2010}, and relatively few absorption lines sensitive to chromospheric activity \citep{Vern1981}. However, the chromosphere demands an order of magnitude greater energy flux than the corona to maintain a temperature of $\sim10{\,}000$~K \citep{Withbroe1977}, placing a greater importance on identifying the heating processes at these lower heights. 

In the denser and partially ionised chromosphere, large-scale current sheet formation and flare activity have not been shown to effectively heat localised plasma \citep{Socas2005}. Instead, the tenative observational evidence of nanoflares presents the only postulated mechanism for chromospheric flare heating \citep{Jess2014, Jess2019, 2018ApJ...862L..24P}. In contrast, there is an abundance of magnetohydrodynamic (MHD) wave observations \citep[see the reviews of][]{Jess2015, 2016GMS...216..431V, 2021JGRA..12629097S} to corroborate the previously proposed dissipation of waves generated at the solar surface into localised chromospheric plasma \citep[e.g.,][]{Sch1948, Leigh1962, Noyes1963}. In subsequent studies of wave generation and propagation in the lower atmosphere, it became clear that despite oscillations being generated across the entire solar surface, the majority of which are in the $3-5$ minute $p$-mode periodicity range \citep{Braun1988}, much of this acoustic energy flux could not penetrate into the chromosphere, either through wave reflection at the chromospheric boundary, or shock formation at lower heights \citep{Narain1996, Fossum2005}. Rather, the strong, vertical magnetic fields of active regions can guide global resonant MHD wave modes into the upper regions of the atmosphere to influence heating \citep{Bel1977, Khom2015}.

Magnetic pores are often considered as the pre-cursor, or aftermath to, the formation of a sunspot \citep{Garcia1987}. Their smaller size and field strengths in comparison to sunspots make them more dynamic and reactive to wave generation mechanisms, such as convective buffeting \citep{Sob2003}, whilst having lifetimes far exceeding other smaller magnetic flux tubes \citep{Keys2014}. As a result, there has been a plethora of modern observations of various MHD wave modes within pores \citep[e.g.,][]{Centeno2009, Stan2011, Stan2012, Cho2015}. In particular, compressible sausage mode waves have been prevalent in the lower atmosphere since their first detection \citep{Doro2008}, distinguished from other modes through complementary intensity and cross-sectional area oscillations in the pore \citep{Edwin1983, Morton2011, Moreels2013b}. Sausage mode signatures have confirmed the viability of pores as wave guides across the solar surface, where \citet{Keys2018} observed both surface and body modes to be ubiquitous within a large sample of photospheric pores. These compressible motions within pores have also been seen to propagate into the chromosphere, with extensive wave damping along their direction of propagation of approximately $50{\,}000$~W/m$^{2}$ between the photospheric and chromospheric layers \citep{Grant2015, Moreels2015}. 

Recently, \citet[][henceforth referred to as \citetalias{GM21}]{GM21} employed inversions of spectropolarimetric Si~{\sc{i}}~10827{\,}{\AA} data to further constrain sausage mode damping in a series of adjacent pores, with further identification of extensive wave damping in the range of $25{\,}000-30{\,}000$~W/m$^{2}$ in the lower solar atmosphere. Subsequent modelling of this damping was conducted by \citet{Riedl2021}, proposing that localised drivers in the pore were generating the sausage modes, which were damped as a result of wave leakage from the body of the pore, alongside geometric effects from the attenuation of the magnetic field as a function of height. In this study, complementary imaging data to the products used by \citetalias{GM21} is used to infer the nature of the photospheric wave driver in order to confirm the proposed damping mechanisms of \citet{Riedl2021}. Further, the nature of the observed waves as they bridge into the chromosphere is assessed alongside whether the properties of each pore has an effect on wave propagation, using chromospheric spectral data that is characterised further with future solar missions in mind.

\section{Observations and Data Processing}
The data presented here is an observational sequence obtained during $14$:$09-15$:$59$~UT on 2016 July 12 with the Dunn Solar Telescope (DST) at Sacramento Peak, New Mexico. The telescope was focussed on the active region NOAA $12564$, positioned at heliocentric co-ordinates ($-425${\arcsec}, $98${\arcsec}), or N$10.4$E$27.5$ in the conventional heliographic co-ordinate system (see Figure~\ref{ROSA_fov}).

\begin{figure}[t!]
  \centering
  \includegraphics[trim=18mm 5mm 10mm 10mm, clip, width=0.8\columnwidth, angle=0]{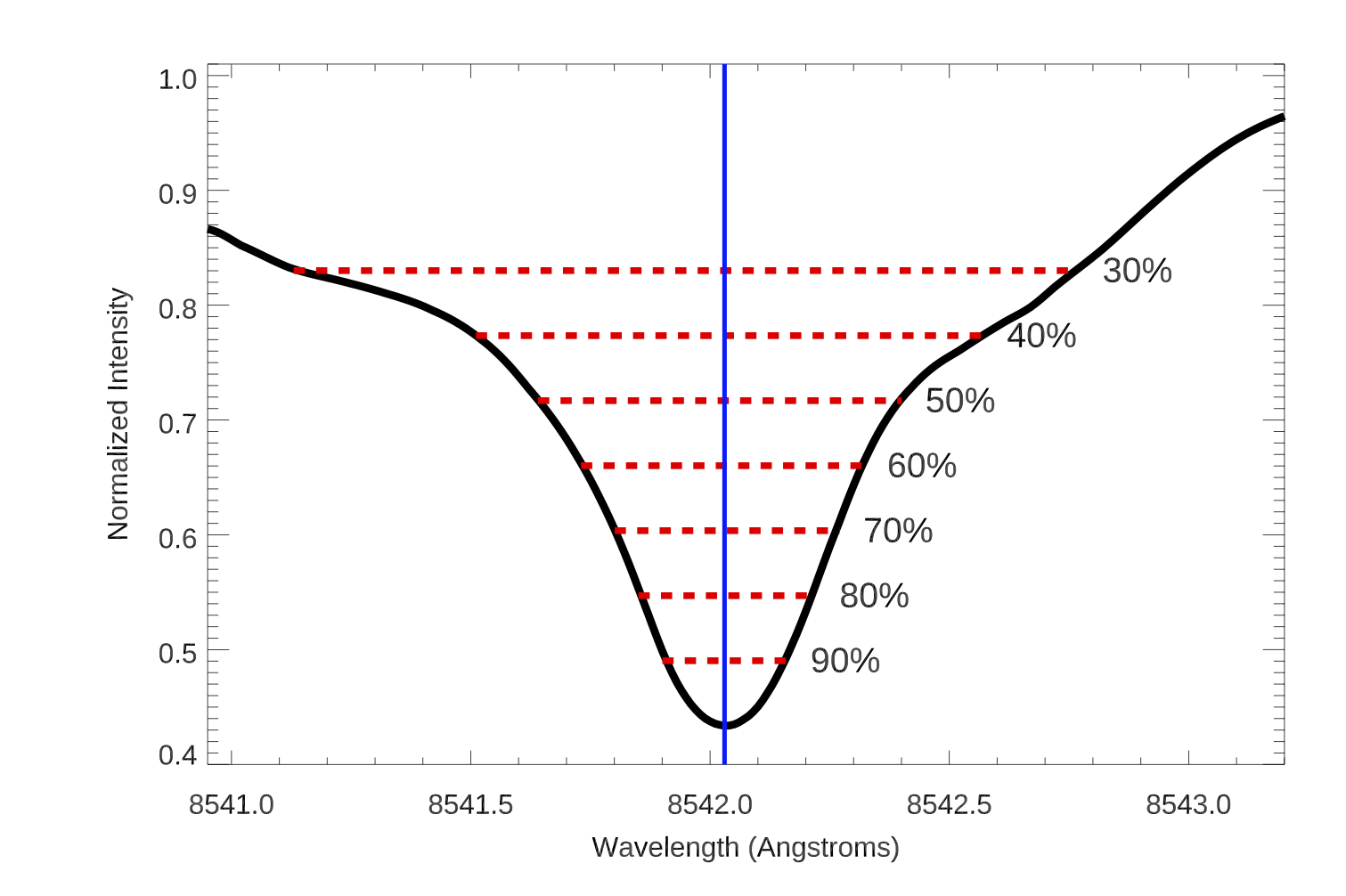}
\caption{The temporally and spatially averaged \CaIR profile for the full field-of-view (including pores, small-scale magnetic elements, and quiet Sun locations). The blue line illustrates the calculated line core for these observations (8542.03{\,}{\AA}), with the red dashed lines highlighting the percentage line depths used to calculate the corresponding bi-sector velocities.}
\label{IBIS_profile}
\end{figure}

The Interferometric BIdimensional Spectrometer \citep[IBIS;][]{Cav2006} was utilized to sample the Ca~{\sc{ii}} absorption profile at 8542.12{\,}{\AA} with 47 non-equidistant wavelength steps employed covering $\pm1.3${\,}{\AA} from the line core (see Figure~\ref{IBIS_profile}). The IBIS instrument imaged a 97{$''$} $\times$ 97{$''$} (approximately $70 \times 70$~Mm$^{2}$) region of the solar disk, with a spatial sampling of $0{\,}.{\!\!}{''}098$ (71{\,}km) per pixel and temporal cadence of 9.4{\,}s per full scan (see Figure~\ref{IBIS_fov}). A whitelight camera, synchronized with the narrowband channel, was also utilized to further correct for seeing effects in the narrowband images. The narrowband images were destretched using co-spatial and co-temporal vectors calculated from a dense grid of sub-field kernels applied to the whitelight image sequence \citep[following the methodology applied in][]{Jess2007, 2010ApJ...719L.134J, Grant2018}. Fourier analysis of the destretch vectors confirmed that such corrections did not add spurious oscillatory power at any frequency under consideration in this study. A blueshift correction was also applied to all narrowband images to account for the use of classical etalon mountings \citep{Cauzzi2008}.

The Rapid Oscillations in the Solar Atmosphere \citep[ROSA;][]{Jess2010} camera system was used to image a $145\arcsec\times145\arcsec$ portion of the solar disk through a continuum ($4170${\,}{\AA}) filter, at a spatial sampling of $0{\,}.{\!\!}{\arcsec}155$ per pixel (see Figure~\ref{ROSA_fov}). The image clarity from both instruments were improved using techniques of high-order adaptive optics \citep{Rimmele2004} and Fourier co-alignment \citep{Jess2007}. For the ROSA imaging data, speckle reconstruction \citep{Woger2008} was also applied. Utilising $64 \rightarrow 1$ speckle restorations, the resulting cadence for the continuum image sequence was $2.11${\,}s.

The seeing remained good for the initial phase of observing, but deteriorated towards the end of the time series, in particular for the $4170${\,}{\AA} continuum ROSA images that were obtained in the near-UV portion of the electromagnetic spectrum. Given the importance of accurately tracking the two-dimensional pore boundaries, only the first 62 minutes of data is used in our study, equating to $400$ IBIS scans and $1800$ ROSA images.

%%%%%%%%%%%%%%%%%%%%%%%%%%%%%%%%%%%%%%%%%%%%%%%
% FIGURE 2
%%%%%%%%%%%%%%%%%%%%%%%%%%%%%%%%%%%%%%%%%%%%%%%
%%%%%%%%%%%%%%%%%%%%%%%%%%%%%%%%%%%%%%%%%%%%%%%
%%%%%%%%%%%%%%%%%%%%%%%%%%%%%%%%%%%%%%%%%%%%%%%
%%%%%%%%%%%%%%%%%%%%%%%%%%%%%%%%%%%%%%%%%%%%%%%

\section{Analysis and Discussion}
Within the field-of-view, a set of magnetic pores are present, formed as a result of the decay and break up of a sunspot. The rate of magnetic field dispersion in this active region is significant, and thus smaller magnetic flux concentrations disappear over the observing period. Five larger magnetic pores are identified throughout the observing time in \citetalias{GM21}, the properties of which are presented in Table~\ref{Pore Properties}, as a result of the Stokes Inversion based on Response functions \citep[SIR;][]{Ruiz1992} analysis of the photospheric Si~{\sc{i}}~10827{\,}{\AA} line (except the area which is determined by intensity thresholding; see below). It is evident that these pores present in a unique configuration, worthy of study at higher spatial and temporal resolution using complementary data products. Initial importance is placed on the interaction of these pores in the low photosphere, and whether they act as individual flux tubes or retain the monolithic behaviour of the preceding sunspot \citep[e.g., consistent with the coherent umbral dynamics presented in][]{2012ApJ...757..160J, 2016NatPh..12..179J, 2017ApJ...842...59J, 2014ApJ...792...41Y, 2021RSPTA.37900181A, 2021A&A...649A.169S}.

\subsection{Photospheric Interactions}
\label{photo}

%%%%%%%%%%%%%%%%%%%%%%%%%%%%%%%%%%%%%%%%%%%%%%%
% FIGURE 3 
%%%%%%%%%%%%%%%%%%%%%%%%%%%%%%%%%%%%%%%%%%%%%%%
\begin{figure*}[t!]
  \centering
  \includegraphics[trim=12mm 0mm 8mm 0mm, clip, width=0.9\textwidth, angle=0]{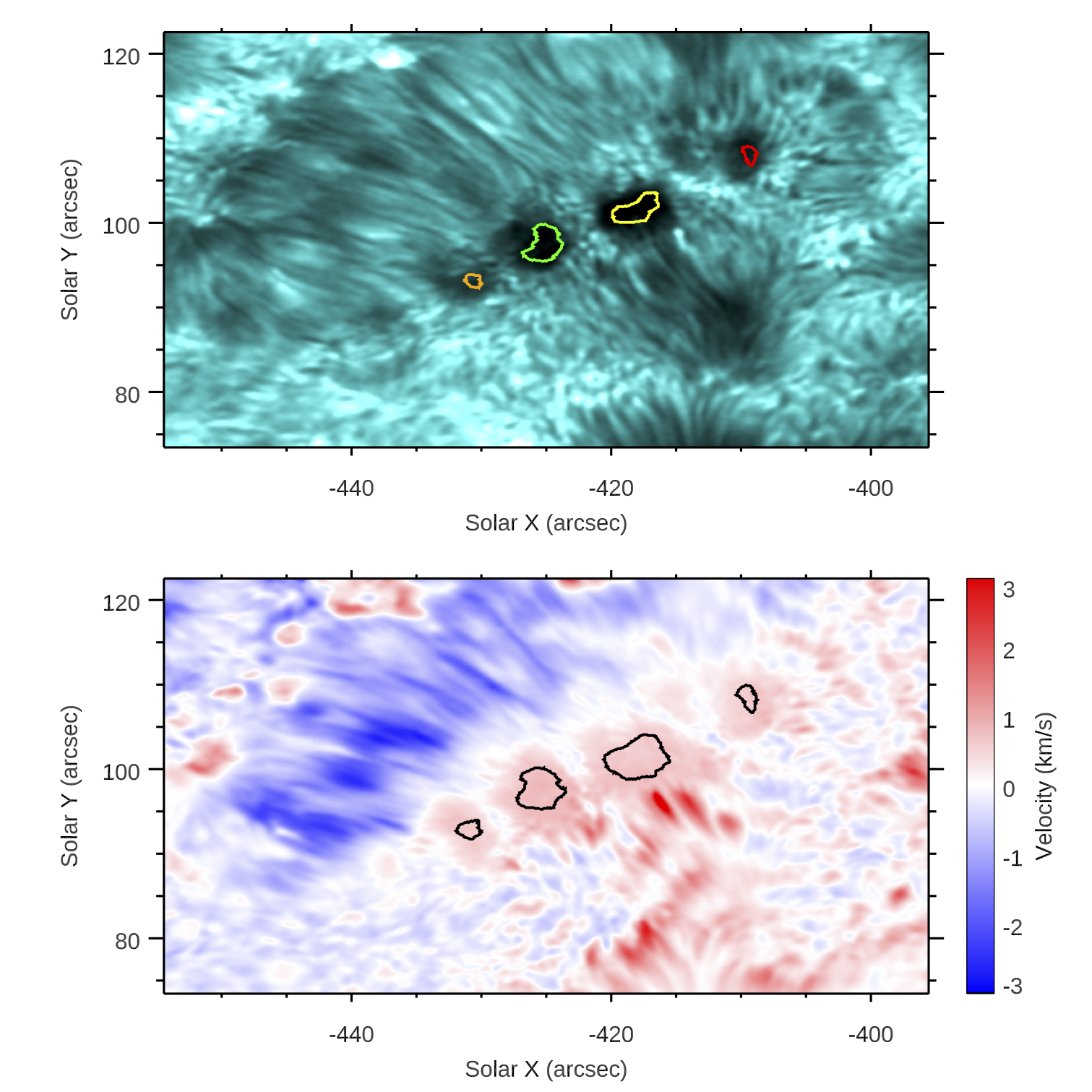}
\caption{The upper panel displays a typical IBIS line core (8542.03{\,}{\AA}) image of the pores in true heliocentric coordinates. The time-averaged perimeters of the established pore boundaries across the bi-sector range are contoured in the same colour scheme as Figure~\ref{ROSA_fov}. The lower panel shows the average line-core line-of-sight velocity of the same region, where a positive value is a plasma down flow (i.e., red-shift), with the time-averaged pore perimeters at line-core contoured in black.}
\label{IBIS_fov}
\end{figure*}
%%%%%%%%%%%%%%%%%%%%%%%%%%%%%%%%%%%%%%%%%%%%%%%
%%%%%%%%%%%%%%%%%%%%%%%%%%%%%%%%%%%%%%%%%%%%%%%
%%%%%%%%%%%%%%%%%%%%%%%%%%%%%%%%%%%%%%%%%%%%%%%

The $4170${\,}{\AA} ROSA continuum filter, which images a height of $\sim25$~km above the solar surface \citep{Jess2012}, was utilised to assess the pores as they become visible in the solar atmosphere. Intensity thresholding was applied to isolate the pores in the image set, in a similar manner to \citet{Grant2015}. A quiescent region, free of any extraneous magnetic brightenings (the green region outlined in Figure~\ref{ROSA_fov}) was used to derive a characteristic mean intensity, $I_{\mathrm{mean}}$, and standard deviation, $\sigma$, at each time step. The pores were then identified individually as significant clusters of pixels within the reduced field-of-view exhibiting less intensity than a threshold of $I_{\mathrm{mean}} - 2.5\sigma$. Due to the dynamic evolution of small-scale magnetic flux in the active region, a minimum size threshold of 50 pixels ($\sim630{\,}000$~km$^{2}$) was placed on identified pixel groupings, with any objects below this excluded. This had the effect of under-representing the perimeter of pore 5 on occasion, as it branches into two objects for a short period due to convective buffeting. However, it is of greater importance to retain only the pixels that definitely constitute the five pores in order to ensure any detected oscillations are being channeled by the larger flux tubes (i.e., minimising any wave contributions from neighbouring non-pore photospheric plasma). The average areas of the pores (henceforth referred to as P1 -- P5; see Figure~{\ref{ROSA_fov}} for their specific labels) were calculated and are presented in Table~\ref{Pore Properties}, confirming that the active region was configured as two large central pores, P3 \& P4, bounded by smaller pores, P1, P2 \& P5.

%%%%%%%%%%%%%%%%%%%%%%%%%%%%%%%%%%%%%%%%%%%%%%%
% TABLE
%%%%%%%%%%%%%%%%%%%%%%%%%%%%%%%%%%%%%%%%%%%%%%%
\begin{table*}[ht]
\caption{Properties of the five pores identified in GM21 at a height of $\sim 25$~km}
\label{Pore Properties}      % is used to refer this table in the text
 \centering
\begin{tabular}{l c c c c c}
\hline\hline
Parameters & {Pore 1} & {Pore 2} & {Pore 3} & {Pore 4} & {Pore 5} \\    
\hline                        % inserts single horizontal line
Magnetic Field (kG) & $1.49 \pm 0.12$ & $1.58 \pm 0.16$ & $1.68 \pm 0.19$ & $1.68 \pm 0.21$ & $1.49 \pm 0.11$ \\ 
Temperature (kK) & $5.67 \pm 0.07$ & $5.34 \pm 0.28$ & $5.18 \pm 0.28$ & $5.20 \pm 0.23$ & $5.48 \pm 0.14$\\
Log$_{10}$~Density (kg/m$^{3}$) & $-3.19 \pm -4.75$ & $-3.15 \pm -4.25$ & $-3.13 \pm -4.24$ & $-3.13 \pm -4.35$ & $-3.17 \pm -4.51$ \\
Area (Mm$^{2}$) & $1.1 \pm 0.03$ & $5.88 \pm 0.8$ & $12.01 \pm 0.8$ & $11.73 \pm 0.7$ & $7.5 \pm 0.3$ \\
\hline
\end{tabular}
\end{table*}
%%%%%%%%%%%%%%%%%%%%%%%%%%%%%%%%%%%%%%%%%%%%%%%
%%%%%%%%%%%%%%%%%%%%%%%%%%%%%%%%%%%%%%%%%%%%%%%
%%%%%%%%%%%%%%%%%%%%%%%%%%%%%%%%%%%%%%%%%%%%%%%

\begin{figure}[t!]
  \centering
  \includegraphics[width=0.8\columnwidth, angle=0]{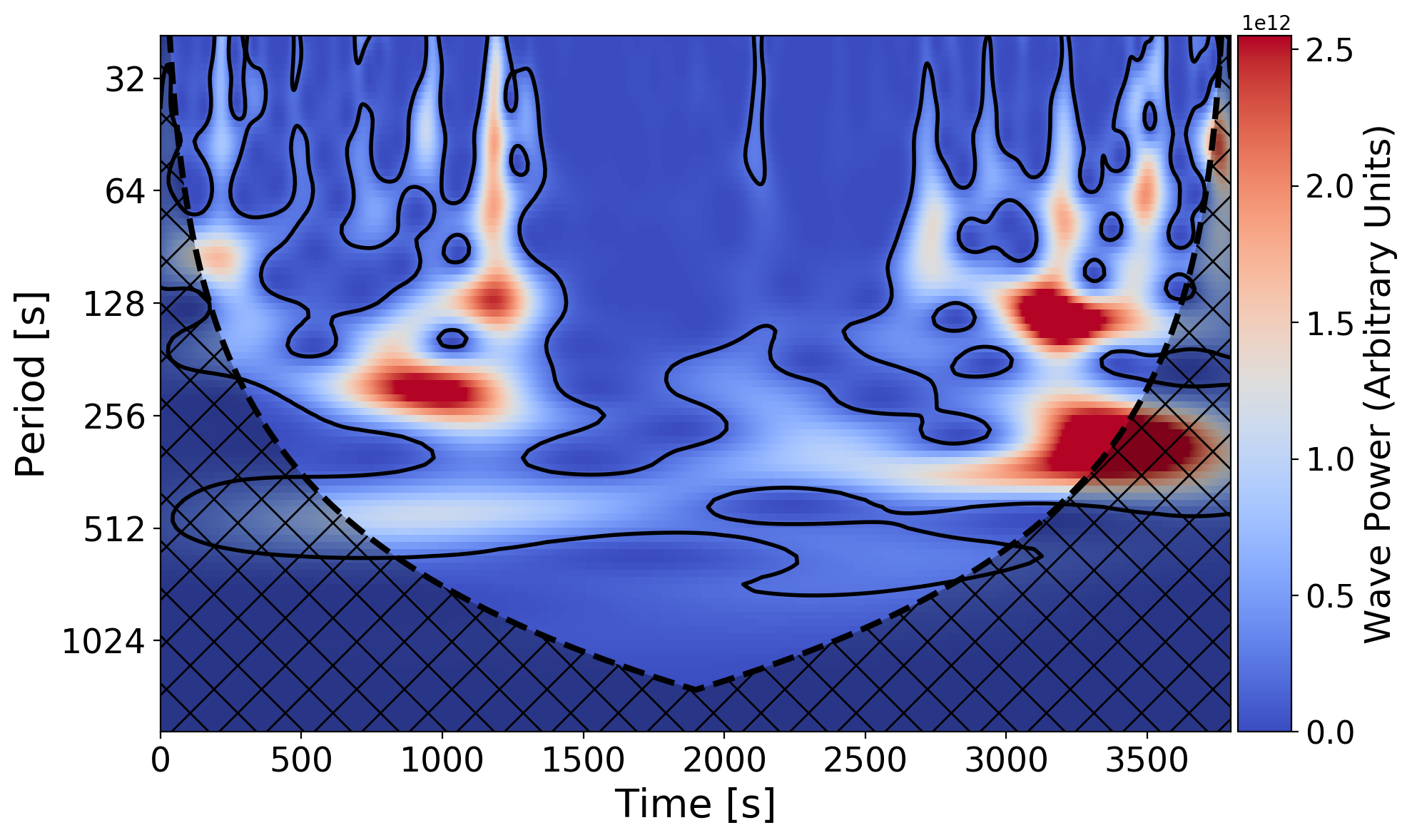}
\caption{The wavelet power spectrum of the cross-sectional area time series of P3 from the $4170${\,}{\AA} ROSA continuum filter. The contoured region are those regions calculated to be above the 95\% confidence of containing coherent oscillatory power. The hatched region to the bottom of the figure represents the cone-of-influence.}
\label{Wavelet}
\end{figure}

The identification of pore boundaries allows for the extraction of plasma characteristics to confirm the existence of sausage-mode oscillations. Following the methodology of \citet{Moreels2013b}, time series of area, Lagrangian (total) intensity, and Eulerian (average) intensity were calculated for each pore. Morlet wavelet analysis \citep{Torrence1998} was utilized to infer information on quasi-periodic signals, with each time series first de-trended using a linear line of best fit and mean normalized. As seen in Figure~\ref{Wavelet}, significant quasi-periodic oscillatory power was found in the $3-5$ minute period band, peaking at $\sim210$~s ($4.76$~mHz) for each pore in both intensity and area, consistent with the solar photospheric $p$-mode spectrum \citep{Lites1982} and previous reporting from \citetalias{GM21}.
%%%%%%%%%%%%%%%%%%%%%%%%%%%%%%%%%%%%%%%%%%%%%%%
% FIGURE WAVELET
%%%%%%%%%%%%%%%%%%%%%%%%%%%%%%%%%%%%%%%%%%%%%%%
%%%%%%%%%%%%%%%%%%%%%%%%%%%%%%%%%%%%%%%%%%%%%%%
%%%%%%%%%%%%%%%%%%%%%%%%%%%%%%%%%%%%%%%%%%%%%%%
%%%%%%%%%%%%%%%%%%%%%%%%%%%%%%%%%%%%%%%%%%%%%%%

%%%%%%%%%%%%%%%%%%%%%%%%%%%%%%%%%%%%%%%%%%%%%%%
% FIGURE 4
%%%%%%%%%%%%%%%%%%%%%%%%%%%%%%%%%%%%%%%%%%%%%%%
\begin{figure}[t!]
  \centering
  \includegraphics[trim=28mm 10mm 10mm 10mm, clip, width=0.7\columnwidth, angle=0]{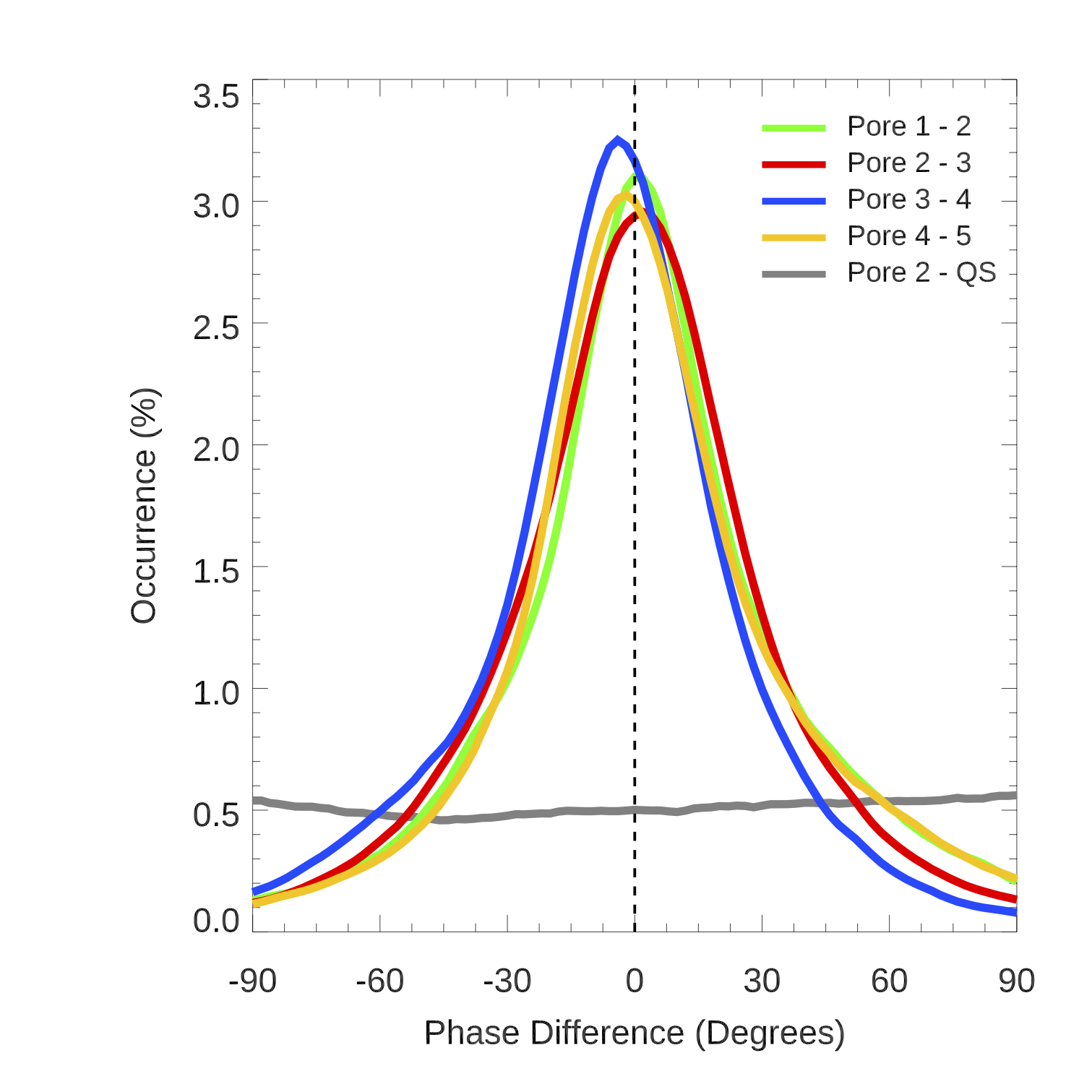}
\caption{Histograms of the phase differences between coherent oscillations within pixels of adjacent pores. The dashed black line represents a phase lag of zero degrees. The histogram of P1 -- P2 is in green, P2 -- P3 in red, P3 -- P4 in blue, P4 -- P5 in yellow, while the comparison between P2 and a quiet Sun region is displayed in grey.}
\label{ROSA_phase}
\end{figure}
%%%%%%%%%%%%%%%%%%%%%%%%%%%%%%%%%%%%%%%%%%%%%%%
%%%%%%%%%%%%%%%%%%%%%%%%%%%%%%%%%%%%%%%%%%%%%%%
%%%%%%%%%%%%%%%%%%%%%%%%%%%%%%%%%%%%%%%%%%%%%%%

%%%%%%%%%%%%%%%%%%%%%%%%%%%%%%%%%%%%%%%%%%%%%%%
% FIGURE WAVELET COMP
%%%%%%%%%%%%%%%%%%%%%%%%%%%%%%%%%%%%%%%%%%%%%%%
\begin{figure}[t!]
  \centering
  \includegraphics[width=0.8\columnwidth, angle=0]{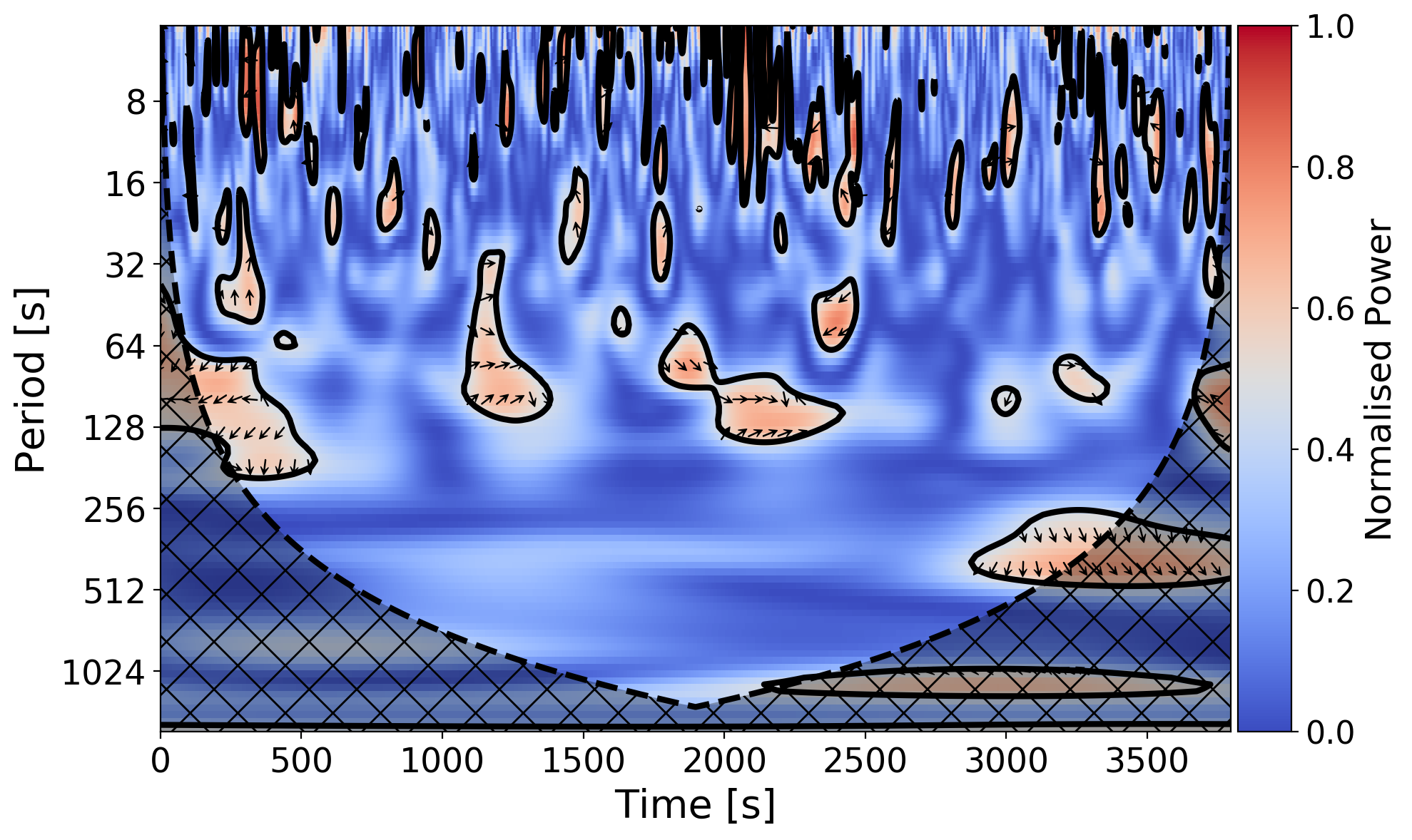}
\caption{The normalized cross power spectrum from the wavelet analysis between the average intensity emission of P3 and the reference magnetic feature. Contoured regions represent the values above the $95\%$ confidence interval, with the phase angle between the time series represented by arrows (a $0\degr$ phase lag is represented by an arrow pointing
right, $180\degr$ by a left arrow, and $90\degr$ and $-90\degr$ by up and down arrows respectively). The cone-of-influence is represented by the hashed region.}
\label{Wavelet_comp}
\end{figure}
%%%%%%%%%%%%%%%%%%%%%%%%%%%%%%%%%%%%%%%%%%%%%%%
%%%%%%%%%%%%%%%%%%%%%%%%%%%%%%%%%%%%%%%%%%%%%%%
%%%%%%%%%%%%%%%%%%%%%%%%%%%%%%%%%%%%%%%%%%%%%%%

The identification of perturbations in the area of the pores denotes compressible sausage modes, the form of which can be confirmed by assessing the phase relationship between area and Lagrangian intensity oscillations \citep{Moreels2013b}. Wavelet phase analysis was utilised \citep{Torrence1999}, with only common oscillations in both time series exhibiting a lifetime greater than $\sqrt{2}{\,}P$, where $P$ is the period of the wave, and a normalised cross-correlation co-efficient greater than $0.5$ considered, encompassing the wave power above the 95\% confidence interval \citep{Torrence1999, Grant2015}. These steps were taken to ensure that detected signals were periodic, and were indeed the same perturbation being considered in both time series. For each pore, a strong in-phase relationship between area and Lagrangian intensity is found, with the relative phases being close to $0\degr$. For each pore, the phase difference between the area and Lagrangian intensity is: P1~${= -2.61 \pm 4.23\degr}$, P2~${= -2.65 \pm 3.07\degr}$, P3~${= -2.21 \pm 2.27\degr}$, P4~${= -0.16 \pm 1.18}$, and P5~${= -1.49 \pm 1.72\degr}$. The strength of this in-phase behaviour confirms the existence of slow-mode sausage waves in every magnetic structure \citep{Moreels2013b}, corroborating the conclusions of \citetalias{GM21}, and further supporting the ubiquity of sausage modes in photospheric pores \citep{Keys2018}.

The classification of compressible MHD waves in each pore naturally leads to the study of whether these modes are exclusive to each structure, or exhibit a coherency across the active region. This was initially conducted through wavelet phase analysis of the area, Lagrangian, and Eulerian intensity time series between adjacent pores, using the same criteria previously determined. The phase difference between the structural oscillations of the pores are documented in Table~\ref{total phases}, and reveal a definite commonality between the pores, with every oscillatory property exhibiting phases centred around zero degrees. This observed coherence is indicative of a common wave driver acting on each structure in unison, and provides a novel configuration where waves with equivalent initial properties can be studied across differing magnetic structures. In order to confirm such an assertion, further analysis was necessary to assess the coherency of perturbations across the entire field-of-view. 

%%%%%%%%%%%%%%%%%%%%%%%%%%%%%%%%%%%%%%%%%%%%%%%
% TABLE
%%%%%%%%%%%%%%%%%%%%%%%%%%%%%%%%%%%%%%%%%%%%%%%
\begin{table*}[t!]
\caption{The phase relationships of oscillatory properties between adjacent pores.}
\label{total phases}      % is used to refer this table in the text
 \centering
\begin{tabular}{l c c c}
\hline\hline
Pore Comparison~~~~ & {~~~Area ($\degr$)~~~} & {~~~Lagrangian ($\degr$)~~~} & {~~~Eulerian ($\degr$)~~~} \\ 
\hline                       
Pore 1 -- 2 & $1.05 \pm 2.71$ & $0.95 \pm 1.51$ & $-1.09 \pm 1.66$  \\ 
Pore 2 -- 3 & $-1.14 \pm 2.43$ & $-1.05 \pm 1.27$ & $-0.22 \pm 1.21$ \\ 
Pore 3 -- 4 & $-2.35 \pm 2.56$ & $0.06 \pm 0.88$ & $2.67 \pm 0.89$ \\
Pore 4 -- 5 & $3.72 \pm 2.59$ & $1.52 \pm 1.32$ & $-0.42 \pm 0.67$ \\
Pore 1 -- 5 & $-0.59 \pm 3.45$ & $1.27 \pm 1.61$ & $0.86 \pm 1.12$\\
\hline
\end{tabular}
\end{table*}
%%%%%%%%%%%%%%%%%%%%%%%%%%%%%%%%%%%%%%%%%%%%%%%
%%%%%%%%%%%%%%%%%%%%%%%%%%%%%%%%%%%%%%%%%%%%%%%
%%%%%%%%%%%%%%%%%%%%%%%%%%%%%%%%%%%%%%%%%%%%%%%

First, a statistical study of the relationship between intensity oscillations in neighbouring pores was undertaken on a pixel-by-pixel basis. In order to extract a large sample of representative intensity time series from within the pore boundaries for analysis, the identification of pixels corresponding to locations that exist within the time-dependent pore perimeters at all times was necessary. This was achieved through the co-addition of binary maps of the pores across all times, where in each imaging frame the pore structures were given a value of `1', and all other locations were set to `0'. Once normalised by the number of images, the locations equal to 1 were extracted for investigation. Due to the morphology of the pores throughout the observing time in this decaying active region, the subsets of `constant' pore pixels represent fractions of the total pore areas, with the natural exclusion of pixels at the perimeters of the pores, which are the most impacted by any shifts introduced by seeing effects or the techniques used to correct them. For each of the pores, the number of pixels in this subset, alongside the value quantified as the percentage of the total pixels that fall within their boundaries are as follows: P1~${= 22}$ (25\%), P2~${= 198}$ (43\%), P3~${= 472}$ (50\%), P4~${= 379}$ (41\%), and P5~${= 126}$ (21\%). 

The coherency of adjacent pore oscillations was probed through further wavelet phase analysis, whereby every pixel in a pore was analysed with respect to every pixel in its adjacent pore. This provided a large sample of distinct phase calculations, for instance, in the case of P3 -- P4, $178{\,}888$ individual phase relationships were calculated. Every phase angle value that corresponded to the previously outlined criteria were then collected, with $\sim10^{8}$ extracted phase results in each case. The results of each study are visualised as histograms in Figure~\ref{ROSA_phase}, where they display a Gaussian-like distribution, with clear dominant peaks around zero degrees, with the most prominent phase differences manifesting at: P1--P2~${= 1.08\degr}$, P2--P3~${= 1.67\degr}$, P3--P4~${= -5.45\degr}$, and P4--P5~${= -5.62\degr}$. The direct comparison of distinct regions (e.g., a central and perimeter pixel) in adjacent pores leads to a larger range of observed phase angles, where small deviations away from zero degrees due to compositional differences in the plasma column (affecting the associated optical depths), in addition to the Wilson effect \citep{Wilson1968}, are likely. Despite this, the coherence of waves across the pores has been corroborated by the clear display of similarly shaped distributions, alongside associated dominant phase lags approaching zero degrees. 

Further analysis was necessary to ensure the distributions in Figure~\ref{ROSA_phase} were neither a statistical effect, spurious information added through instrumental noise, such as the uncorrected tail of the point spread function (PSF) of the telescope \citep{2006SPIE.6272E..3WM} or atmospheric turbulence \citep{Rimmele2004}. Two processes were undertaken to verify the derived relationships of in Figure~\ref{ROSA_phase}. Firstly, wavelet phase analysis was conducted between the pixels in P2 and the quiet-Sun region outlined by the green box in Figure~\ref{ROSA_fov}. The consideration of a region free of strong magnetic flux and distant from the active region will confirm the uniqueness of the commonality of waves in the pores, with the influence of the PSF limited to its full width at half maximum and the isoplanatic patch caused by turbulence tending to $\sim5-10$~arcsec. Common periodicities of the embedded wave activity is found between the pore and quiet Sun region, as to be expected as a result of ubiquitous photospheric $p$-mode generation. Critically, the resulting phase distribution, visualized by the grey line in Figure~\ref{ROSA_phase}, shows no preferred phase lag, with all values being approximately likely. 
The lack of any discernible relationship between these regions infers that the commonality of the pores is not due to any instrumental effects. However, it is also prudent to confirm that the pore relationship is not influenced by seeing effects, such as straylight enhancements of dark regions, or the reconstruction techniques used to correct for them. This was conducted through a comparison of P3 with a small magnetic structure separate to the cluster of pores (as seen in Figure~\ref{ROSA_fov}). This feature is noticeably smaller than the pores under study, with its perimeter contoured by a threshold of $I_{\mathrm{mean}} - 1.5\sigma$. This made the object more transient in nature, with significant morphological and location changes across the observing window. As a result, no pixels that contained consistent magnetic plasma throughout the time series could be identified, prohibiting a statistical study as seen in Figure~\ref{ROSA_phase}. Instead, wavelet analysis of the average emergent intensity of this structure and P3 was conducted, and is displayed in Figure~\ref{Wavelet_comp}. From inspection, there are limited points in the time series where the two structures share common oscillations, and mostly at higher frequencies than those associated with global $p$-modes. It is also clear that those correlated periodicities are not in-phase, with an average phase angle of ${-93.8 \pm 104.2\degr}$ from those regions above the $95\%$ confidence interval, with the larger range of observed values likely being influenced by the introduction of perimeter pixel measurements and their associated uncertainties as discussed above. The comparison with a separate magnetic structure has shown that there are negligible effects on the coherent nature of the waves in the pores due to any effects across the field of view. It has also shown that the tails of the distributions in Figure~\ref{ROSA_phase} are likely caused by the high-frequency, anti-phase spurious signals seen in Figure~\ref{Wavelet_comp}, strengthening the validity of the peaks around zero degrees. These analysis steps further verify that the negligible phase lags observed between adjacent pores are a real effect, and confirms the commonality of the wave driver within the 5 pore structures.   

%%%%%%%%%%%%%%%%%%%%%%%%%%%%%%%%%%%%%%%%%%%%%%%
% FIGURE 5
%%%%%%%%%%%%%%%%%%%%%%%%%%%%%%%%%%%%%%%%%%%%%%%
\begin{figure*}[t!]
  \centering
   \includegraphics[width=\textwidth, angle=0]{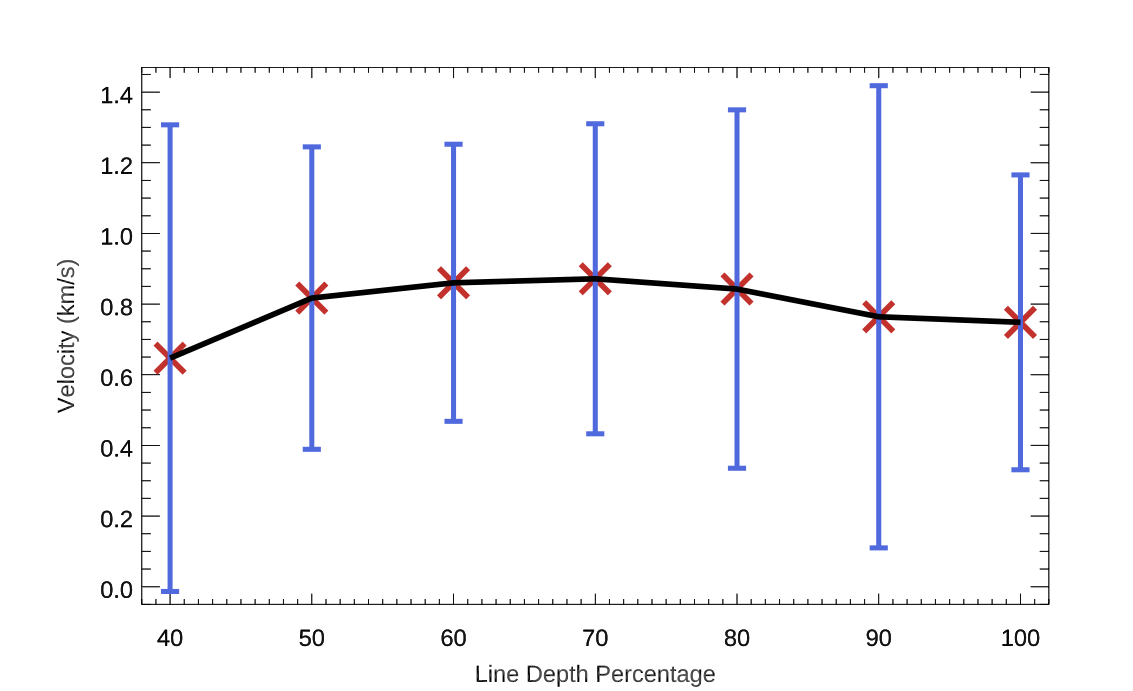}
  \includegraphics[trim=11mm 11mm 11mm 11mm, clip, width=0.49\textwidth, angle=0]{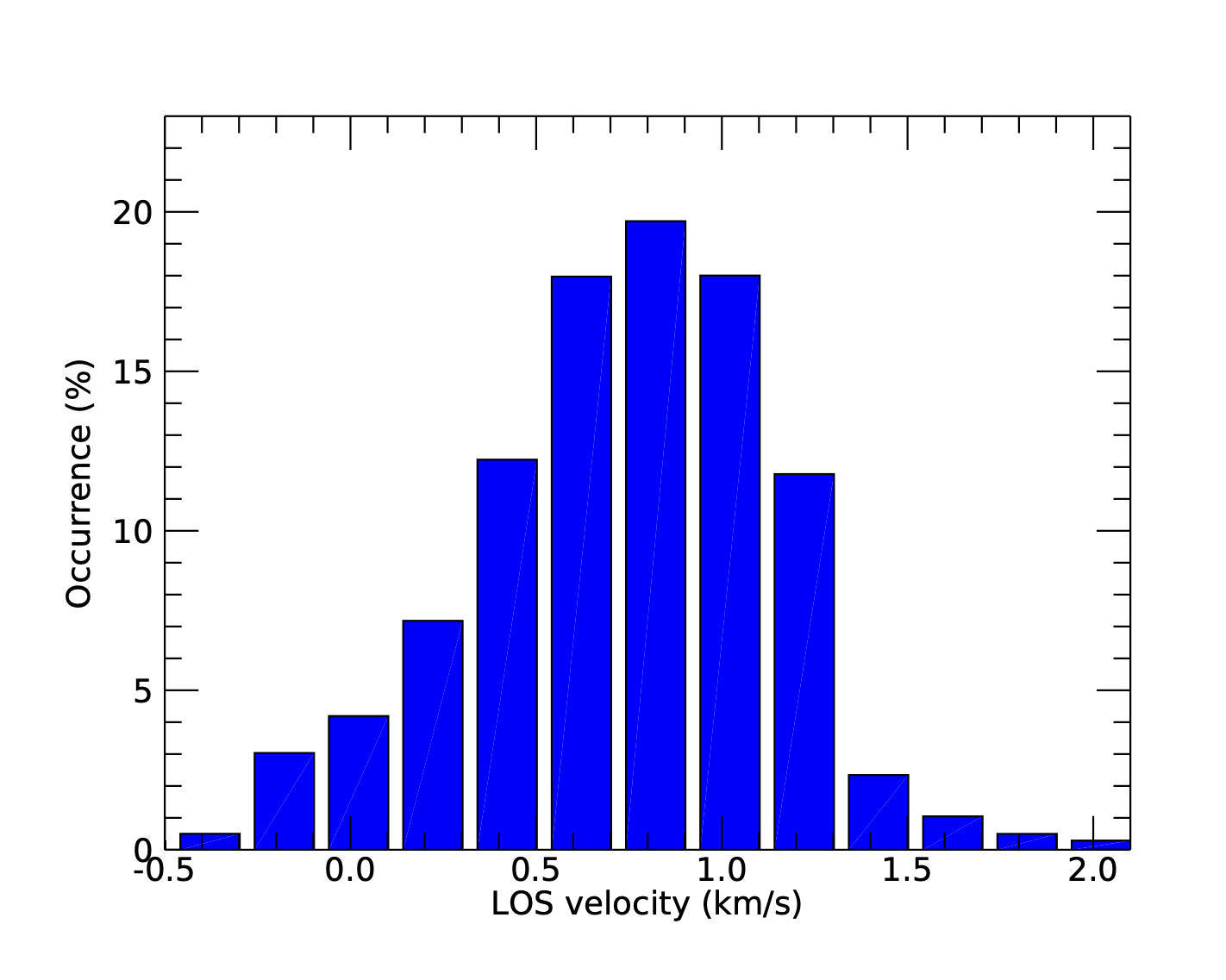}
    \includegraphics[trim=11mm 11mm 11mm 11mm, clip, width=0.49\textwidth, angle=0]{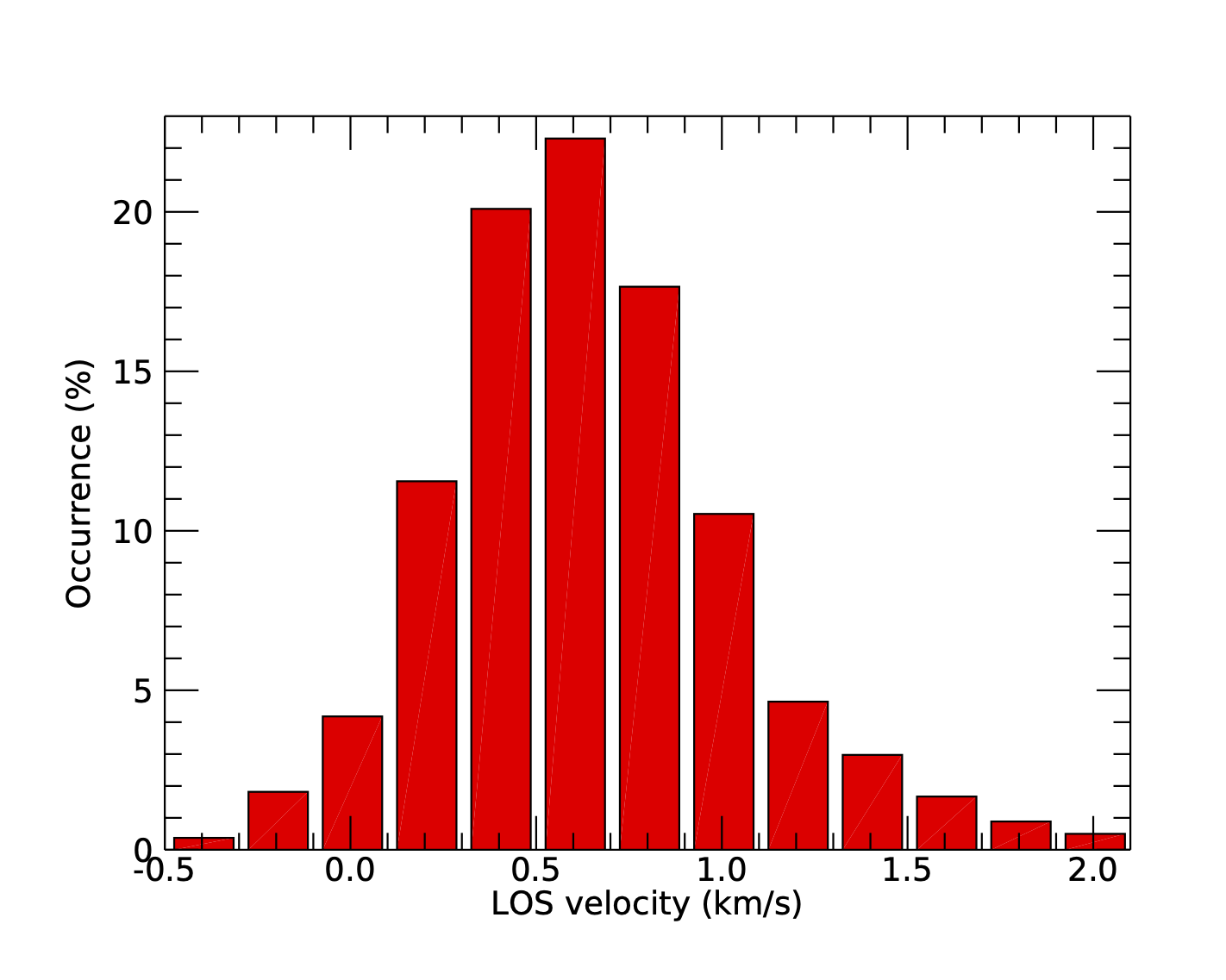}
\caption{Top panel: The average LOS velocity for P3 at each percentage line depth, with standard deviations plotted for each measurement. Bottom panels: Histograms of the LOS velocities of the pixels contained within the P3 contour spanning the full height range of measurements. The left panel (blue bars) represents the velocities at 50\% line depth, while the right panel (red bars) displays the line core velocities calculated at 100\% line depth.}
\label{bisector_velocities}
\end{figure*}
%%%%%%%%%%%%%%%%%%%%%%%%%%%%%%%%%%%%%%%%%%%%%%%
%%%%%%%%%%%%%%%%%%%%%%%%%%%%%%%%%%%%%%%%%%%%%%%
%%%%%%%%%%%%%%%%%%%%%%%%%%%%%%%%%%%%%%%%%%%%%%%

The ubiquitous $p$-mode waves seen across the photosphere are generated in the sub-surface convection zone, through the entropy fluctuations and Reynolds stresses associated with the mixing of magnetic fields and turbulent plasma \citep{Nord2001, Stein2001}. The excitation of $p$-modes appears to be most prominent just below the interface between the convection zone and the solar surface, where the perturbations in convective plasma parameters are largest \citep{Nord2009}. The driver itself can manifest as either a localised force, or as an extended force across a large plane. The observations presented here would indicate the latter, given the coherence of waves across multiple pores spanning tens of Mm. However, the results of \citet{Riedl2021} offer a contrasting view. In their two-dimensional simulations of P3, an extended driver across the lower boundary of both the pore and neighboring quiet-Sun could not replicate the damping seen in \citetalias{GM21}, and in fact produced an increase in wave energy flux with height. Instead, they found good correlation with the observations of \citetalias{GM21} through the application of a localised driver within the pore. This scenario, where each pore has a distinct, localised driver acting upon it to produce equivalent oscillations is unlikely, as the frequency distribution of generated waves in the convection zone is dependent on both the magnetic field strength and inclination of flux tubes \citep{Jacoutot2008, Kiti2011}. Instead, it appears more likely that the waves must be generated at a lower height in the convection zone where the emerging flux bundle is monolithic, before it separates into smaller fragments, manifesting at the surface as the magnetic pores \citep[e.g.,][]{Zwaan1985, Cheung2010, Feng2017}. This scenario allows for each pore in the photosphere to exhibit an identical distribution of MHD oscillations, and thus provides a unique opportunity to study the development of equivalent perturbations in differing wave guides into the chromosphere.

\subsection{Chromospheric Propagation}

The \CaIR line was employed to study the pores as they branch into the chromosphere. This spectral line has been a key chromospheric observable for decades, and will be a key diagnostic for multiple instruments at next-generation observatories such as the Daniel K. Inouye Solar Telescope \citep[DKIST;][]{Rim2020, 2021SoPh..296...70R}. The collection of high cadence spectral scans of this line allowed Doppler velocity signatures to be employed as an additional diagnostic of the pore oscillations. The introduction of Doppler velocity data is necessary, as previous studies have established that chromospheric wave fronts, manifesting as notable line-core intensity brightenings, can propagate across a significant proportion of a pore's surface area at a given time in \CaIR observations \citep[e.g.,][]{Cho2015}. These umbral wave fronts therefore negate time dependent intensity thresholding of the pore perimeters, and prohibit the measurement of chromospheric area perturbations. Instead, the line-of-sight~({\vlos}) velocity of the plasma is inferred for each pixel, with the time-averaged line-core {\vlos} for the field of interest displayed in the lower panel of Figure~\ref{IBIS_fov}. 

When seeking to identify the propagation of MHD waves in the solar atmosphere, multi-line observations are often used in concert to ascertain phase lags \citep[e.g.,][]{Gio1978, Lites1984, Centeno2006}. In this study, the \CaIR spectral line is the only chromospheric observable, thus it is prudent to investigate whether the propagation characteristics of MHD waves can be ascertained from single-line observations of \CaIR through the derivation of bi-sector velocities. 

\subsubsection{\CaIR Bi-sectors}
\label{sec:bi}
%%%%%%%%%%%%%%%%%%%%%%%%%%%%%%%%%%%%%%%%%%%%%%%
% FIGURE 5
%%%%%%%%%%%%%%%%%%%%%%%%%%%%%%%%%%%%%%%%%%%%%%%
\begin{figure*}[t!]
  \centering
  \includegraphics[trim=0mm 7mm 0mm 0mm, clip, width=\textwidth, angle=0]{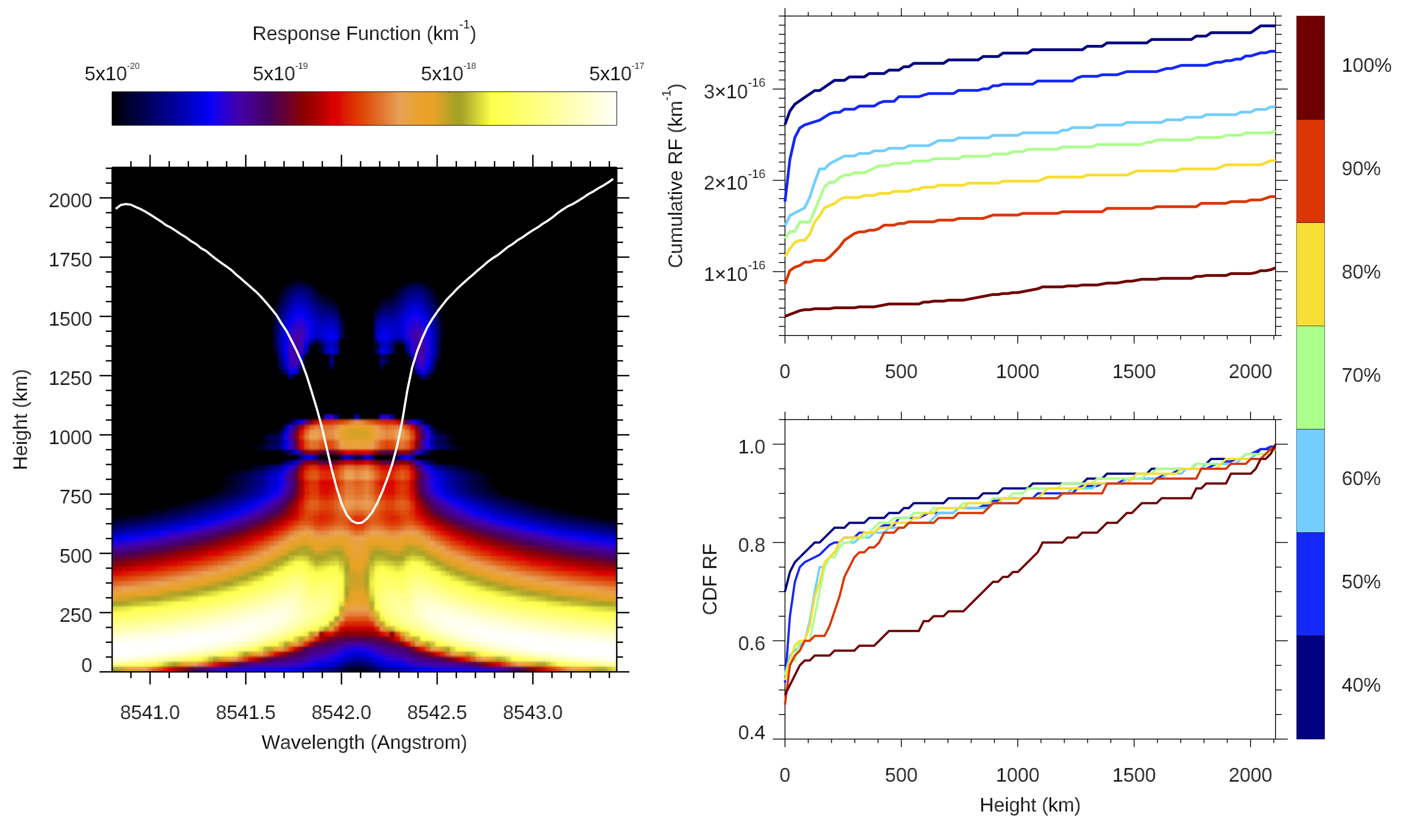}
\caption{Left panel: The response function of the \CaIR spectral line to temperature perturbations in a magnetic atmosphere, with a \CaIR reference profile plotted in white. Upper right panel: The cumulative response function with height, measured at the wavelengths corresponding to each line depth, which are colored according to the legend on the right. Lower right panel: The cumulative distribution function of each bi-sectors response function, with the same color key as above.}
\label{rf}
\end{figure*}
%%%%%%%%%%%%%%%%%%%%%%%%%%%%%%%%%%%%%%%%%%%%%%%
%%%%%%%%%%%%%%%%%%%%%%%%%%%%%%%%%%%%%%%%%%%%%%%
%%%%%%%%%%%%%%%%%%%%%%%%%%%%%%%%%%%%%%%%%%%%%%%

%%%%%%%%%%%%%%%%%%%%%%%%%%%%%%%%%%%%%%%%%%%%%%%
% FIGURE 5
%%%%%%%%%%%%%%%%%%%%%%%%%%%%%%%%%%%%%%%%%%%%%%%
\begin{figure*}[t!]
  \centering
  \includegraphics[trim=0mm 0mm 0mm 0mm, clip, width=\textwidth, angle=0]{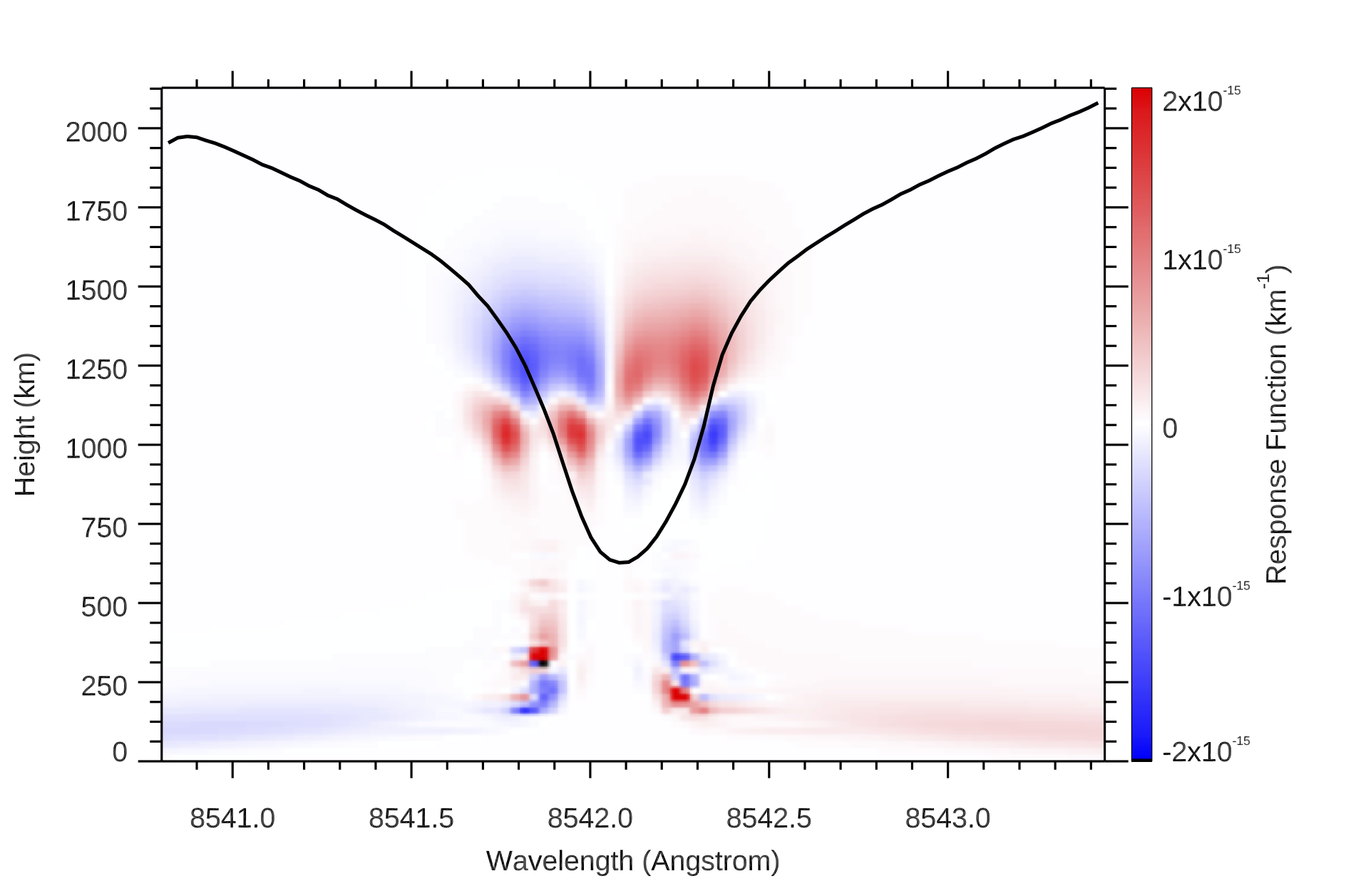}
\caption{The response function of the \CaIR spectral line to a $1$~km s$^{-1}$ velocity perturbation in a magnetic atmosphere, with a \CaIR reference profile plotted in black. Positive (red) values refer to an intensity brightening as a result of the perturbation, with negative (blue) values referring to a reduction in intensity.}
\label{rf_vel}
\end{figure*}
%%%%%%%%%%%%%%%%%%%%%%%%%%%%%%%%%%%%%%%%%%%%%%%
%%%%%%%%%%%%%%%%%%%%%%%%%%%%%%%%%%%%%%%%%%%%%%%
%%%%%%%%%%%%%%%%%%%%%%%%%%%%%%%%%%%%%%%%%%%%%%%

The technique of inferring {\vlos} from various bi-sectors across a spectral line to provide height stratified information is  commonly used for photospheric lines \citep{Kulander1966, 2020A&A...634A..19G}, such as Si~{\sc{i}}~10827{\,}{\AA} in the case of \citetalias{GM21}, due to their symmetrical profiles. However there are only a few examples of inferring \CaIR bi-sector motions in magnetic flux tubes \citep[e.g.,][]{Chae2013, Beck2020}, due to a range of chromospheric effects capable of adding non-linearities to the absorption profile. Therefore it was prudent to first ensure the spectral profiles within the pores were suitable for bi-sector analysis. This was achieved through use of the Multi-Component Atmospheric Line Fitting program \citep[MCALF;][]{Macbride2020, Macbride2021a, MacBride2021b}, in particular the efficient machine learning scheme that classifies \CaIR profiles into five categories based on the degree of line-core emission. \citet{Macbride2021a} highlight that category three (or above) is a result of substantial emission within the line profile, and hence are unsuitable for bi-sector analysis. Thankfully, only classification categories 1 -- 2 were found within or around the vicinity of the pores. This both establishes that the pore observations are suitable for bi-sector analysis, and that none of the wave power detected in the photosphere is being dissipated through detectable shock formation, as the pores lack the associated line-core brightening \citep{Grant2018}. This does not imply that non-linearities are absent in the pores. The atmosphere of a shock contained within an observed pixel has long been established to have two-components, impulsive and quiescent \citep{Socas2000}. The detection of the shock is therefore dependent on the filling factors of these components, and implies that shocks can form below the resolution of current instruments. Given the diversity and complexity of geometric configurations that pores can display \citep[e.g.,][]{Sob2003}, the gradients in density and magnetic field necessary to steepen waves into shocks can manifest in pores. Therefore, any future study of pores in \CaIR must continue to confirm that no discernable emission in the core of the line is present.

Bi-sectors were calculated at locations representing 40 -- 90\% of the line depth relative to the measured average line centre of the data (8542.03{\,}{\AA}), in intervals of 10\% (see Figure~\ref{IBIS_profile}). The resulting bi-sector velocity is a measurement of the shift in wavelength of the \CaIR line wings at these positions, and was conducted for all profiles in the field-of-view. For this data set, only bi-sectors as low as 40\% line depth were used, since isotopic splitting of the \CaIR line produces an asymmetric profile, with a redshifted `inverse-C' shape causing uncertainty at wavelengths far from line centre \citep[i.e., in the 10 -- 30\% line depth interval;][]{Uit2006, Leen2014}. In addition to central reversals from shocks, non-linearities can be introduced to the absorption line through density enhancements in the lower atmosphere. As shown in \citet{Carlsson1997}, the emission source function is coupled to the Planck function in the high photosphere. As a result, density changes in this region produces a notable brightening in the blue wing of the profile, resulting in what would appear to be the red-shifted signature of a downflow \citep[see further discussion in][]{Henriques2020}. The erroneous introduction of a broad redshift due to localised upflows have already been confirmed due to shocks \citep{Henriques2017} and flares \citep{Monson2021}. Despite no signatures of these impulsive events within the pores observed here, incremental density excursions may skew the derived bi-sector velocities, thus inspection of the derived velocities for this effect is necessary.

In order to study the pore oscillations across the derived bi-sector velocities, the chromospheric pore perimeters must be established. The lesser contrast between pores and the quiescent background in the chromosphere necessitated the use of time-averaged images of the field-of-view, in order to better isolate the pore boundaries. In addition, from inspection of the data (see the upper panel of Figure~\ref{IBIS_fov}), it is clear that P1 is not obviously detectable, which is indicative of either the similar chromospheric temperature profiles of small pores to the quiescent background, or the rapid expansion and volume filling of the magnetic fields \citep{Solanki2017}. As a result, pore P1 was excluded from any further study in the \CaIR line. To ensure pore perimeters were valid across every line depth used, a summed wavelength image, spanning $\pm0.54${\,}{\AA} relative to the line core, which encompasses the 40\% line depth region, was generated. The pore thresholding was conducted in an identical manner as described in Section~\ref{photo}, but with bespoke thresholds for each pore, given the variable connectivity between certain pores and dark fibrilar structures. The selected thresholds were: P2 = $I_{\mathrm{mean}} - 2.28\sigma$, P3 = $I_{\mathrm{mean}} - 2.82\sigma$, P4 = $I_{\mathrm{mean}} - 2.84\sigma$, and P5 = $I_{\mathrm{mean}} - 2.04\sigma$, which are used to contour pores P2 -- P5 in the upper panel of Figure~\ref{IBIS_fov}. 

The bi-sector velocities for P3 are plotted in Figure~\ref{bisector_velocities} and are representative of the other pores. It can be seen that there is a small redshifted average of between 0.64 - 0.81~km{\,}s$^{-1}$ (seen in Figure~\ref{IBIS_fov}). However, it is not certain that this is not caused by an upwardly propagating (i.e.,  blueshifted) density enhancement in the lower atmosphere, as chromospheric downflows have also  been associated with rapid cooling within the interior of the pores in the lower atmosphere \citep{Steiner1998, Kato2011, Cho2013}. The lower panels of Figure~\ref{bisector_velocities} show the distribution of velocities on a pixel-by-pixel level for pore P3 at 50\% and 100\% (line-core) line depths. The 50\% line depth was selected to exhibit the full bi-sector range, as from inspection of the top panel the 40\% line depth bi-sector displays a larger variance, likely due to the influence of the previously mentioned `inverse-C' effect. It can be seen that in both samples, there is a predominant trend around the average, and an approximately normal distribution, with excursions at the tail on the order of 1\%. This further reinforces that, for this data set, any skewness as a result of localised density enhancements in the photosphere are minimal, as consistent and expected behaviour is seen within the LOS velocity samples at a range of line depths.

To gain insight into the differences in atmospheric heights sampled by the bi-sectors, the response function of the \CaIR intensities (Stokes~$I$) is considered. The response function describes the sensitivity of emergent intensity to changes in temperature as a function of height and wavelength. The response of the \CaIR line has been analysed for quiescent parts of the solar atmosphere \citep[e.g.,][]{Cauzzi2008}. However, the effects on sensitivity within a strongly magnetic atmosphere has yet to be investigated. To probe the response of the line to pore-like magnetism, the warm sunspot umbral model `L' of \citet{1986ApJ...306..284M} was employed, with the responses of the spectral line calculated according to the method of \citet{2005ApJ...625..556F}. In this method, the atmospheric model is perturbed successively at each height in temperature by 1\% with a step function that is unity from the lower atmospheric boundary to the depth point under consideration, and zero above that. For each perturbation, the change in intensity, $\Delta I_{\lambda}$, is recorded and the response function is derived by calculating the derivative of $\Delta I_{\lambda} / I_{\lambda}$ with atmospheric height. The output of this process is plotted in the left panel of Figure~\ref{rf}, showing the photospheric nature of the \CaIR wings and the higher chromospheric heights sampled towards the line core. 

For reference, the responses of line depth locations for the reference profile in the left panel of Figure~{\ref{rf}} are calculated and visualised as a cumulative response function for each percentage line depth (top right panel of Figure~{\ref{rf}}) and as a cumulative distribution function (CDF) of the total line depth responses (lower right panel of Figure~{\ref{rf}}). The response function shows that the bi-sectors are indeed sampling iteratively higher heights in the solar atmosphere as a function of line depth. However, the CDF shown in the lower right panel of Figure~{\ref{rf}} implies that the bi-sectors still have a considerable contribution from the lower atmosphere, with only wavelengths close to the line core deviating in its development through the atmosphere, providing a notably more purely chromospheric response. 

With the response to emergent intensity derived, the response of the \CaIR line to velocity perturbations can also be calculated. The same atmospheric model is used as depicted in Figure~{\ref{rf}}, however, in this occasion a constant $1$~km{\,}s$^{-1}$ upflow velocity is applied across the atmosphere and subsequently perturbed by 1\%. The resulting intensity change is displayed in Figure~\ref{rf_vel}, with positive response values indicating an intensity increase, while the negative values denote an intensity decrease. From inspection, the effects of a velocity oscillation are predominantly seen in the chromosphere, as expected, and subtend a range of wavelengths around the line core. This implies that there will be a measure of integration of chromospheric velocity signals at these wavelengths. The influence of photospheric velocity perturbations are limited to small portions of the wing, where they will provide a component of integrated response at those wavelengths. However, the strong photospheric contributions seen in Figure~\ref{rf} are not equivalently present in Figure~\ref{rf_vel}, indicating that the large photospheric contributions in the temperature response functions may be over-estimated, especially given the lack of photospheric granulation in the top panel of Figure~\ref{IBIS_fov}. Therefore, it is prudent to test the accuracy of the derived responses, which can be achieved through investigation of pore wave power as a function of frequency throughout the individual line depths. Of particular interest is the cut-off frequency, resulting from the influence of gravity on wave propagation, leading to smaller frequencies being unable to propagate to higher heights \citep{Bel1977}. Various models predict that the cut-off frequency is a height-stratified quantity, with previous attempts made to constrain the atmospheric heights at which these cut-offs manifest \citep[e.g.,][]{Schmitz1998}. However, dependencies on quantities such as magnetic field strength and inclination angles makes it difficult to produce a unified height model \citep{Centeno2009, Felipe2020}. What is clear from observations is the dominance of 5-minute periods in the lower photosphere, to at least heights of $\sim$400~km \citep{Felipe2018, Rajaguru2019}, before these periods are unable to bridge into the chromosphere, allowing 3 minute oscillations to become prominent \citep{Centeno2006, Wis2016}. Therefore, assessment of photospheric contributions to \CaIR bi-sector formation can be examined through the gradual rise of lower frequency wave power as the percentage line depth sampled becomes smaller.

%%%%%%%%%%%%%%%%%%%%%%%%%%%%%%%%%%%%%%%%%%%%%%%
% FIGURE 6
%%%%%%%%%%%%%%%%%%%%%%%%%%%%%%%%%%%%%%%%%%%%%%%
\begin{figure}[t!]
    \centering
  \includegraphics[trim=5mm 8mm 8mm 10mm, clip, width=0.8\columnwidth, angle=0]{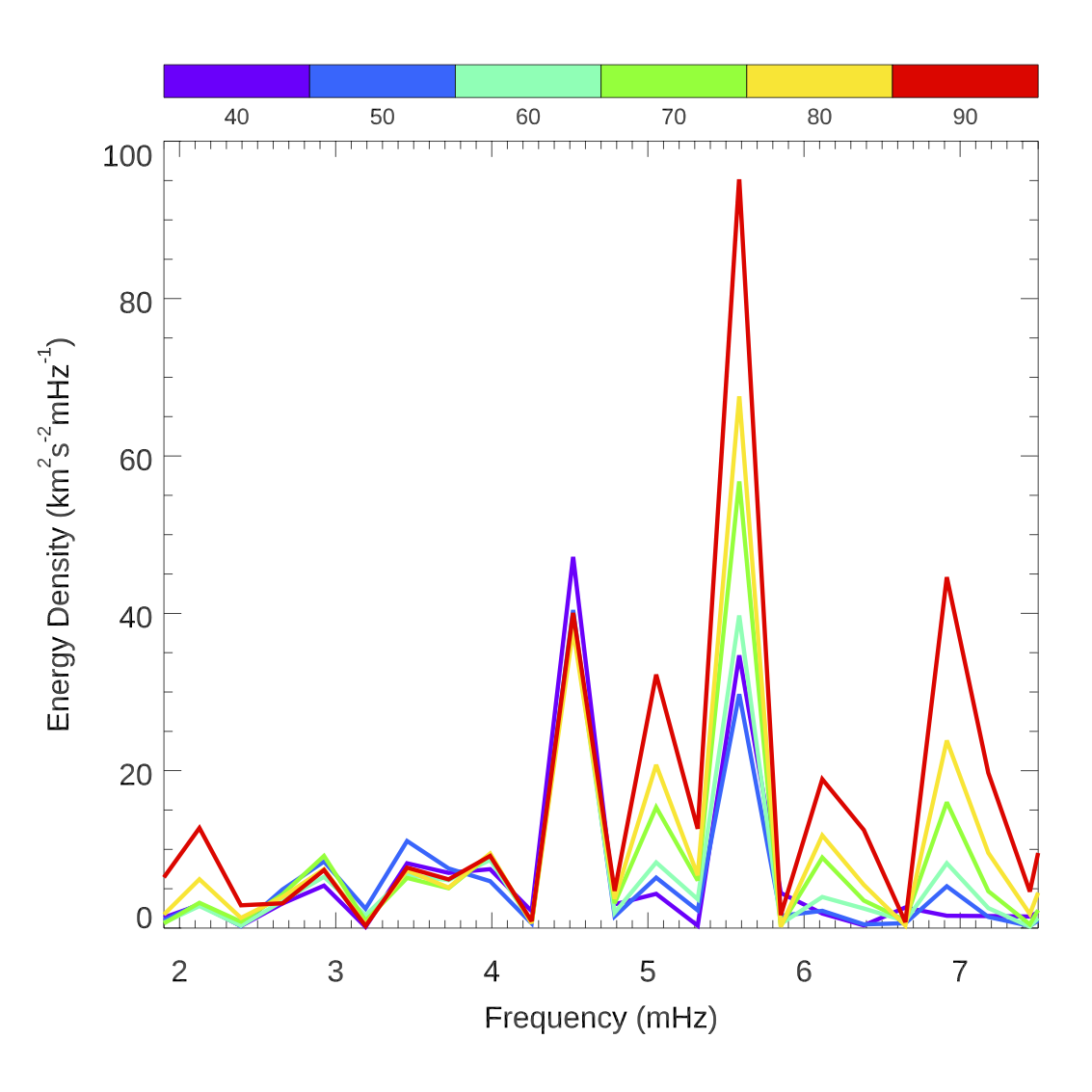}
\caption{The energy spectral density of the spatially-averaged bi-sector velocities from pore P3, plotted in a color scheme represented in the legend above, where the unit corresponds to the percentage line depth of the \CaIR line.}
\label{bisector_density}
\end{figure}
%%%%%%%%%%%%%%%%%%%%%%%%%%%%%%%%%%%%%%%%%%%%%%%
%%%%%%%%%%%%%%%%%%%%%%%%%%%%%%%%%%%%%%%%%%%%%%%
%%%%%%%%%%%%%%%%%%%%%%%%%%%%%%%%%%%%%%%%%%%%%%%

Using the established pore perimeters, the spatially-averaged velocity time series for each of the percentage line depths were isolated for inspection. It was immediately apparent that for all bi-sectors, the 5~mHz (3 minute period) frequency was prominent across all pores. However, Fourier power is a relative quantity, dependent on many factors of the observations, and often normalised to inhibit direct comparison. For a more robust measure of wave activity throughout the bi-sectors, the energy spectral density was calculated for each time-series, where its normalisation relative to the frequency resolution of the time series (i.e., providing a `per mHz' quantity) allows for future direct comparison between different observing regimes. The energy spectral density, $S$, is defined following the convention described by \citet{StullRolandB1988AItB}, 
\begin{equation}
    S = \frac{2|{\bf{X}}(f)|^{2}}{\delta \nu} \ ,
\end{equation}
where ${\bf{X}}(f)$ is the Fourier power spectrum of the time series and $\delta \nu$ is the corresponding frequency sampling. 

The spectral energy densities for pore P3 are shown in Figure~\ref{bisector_density} for inspection. It is clear that the $\sim$3~mHz photospheric contribution is not present in any of the bi-sector velocities, indicating the cut-off height is below that sampled by the 40\% line depth. In contrast, the $\sim$5~mHz chromospheric oscillation is prominent across all 40 -- 90\% line depths, with a shift in the dominant frequency found throughout the line depths, from $\approx$4.5~mHz (221s) at 40\%, to a peak at $\approx$5.6~mHz (179s) for 90\%. As such, the derived bi-sector velocity time series provide observational evidence of the gradual enhancement of three minute wave power as the waves propagate through to higher chromospheric heights. 

In the case of pore P3, the $\approx$5~mHz spectral energy density increases incrementally by a factor of two across the range of 40 -- 90\% of the \CaIR line depth, and in the case of $\approx$7~mHz by a factor of 10. Simulations have highlighted the amplification of higher frequency ($\ge5$~mHz) power as the waves propagate through negative density gradients \citep{Khom2015, Felipe2019}, and although previous observational studies have identified the shift in dominant power \citep[e.g.,][]{Krishna2015}, they lacked the height sampling necessary to track this gradual increase in wave power through the chromosphere. The fidelity of the measured wave enhancement shown in Figure~{\ref{bisector_density}}, allied with the shift in dominant frequency, shows that despite the influence of photospheric heights in the \CaIR response functions (see, e.g., the left panel of Figure~{\ref{rf}}), the bi-sector velocities are capturing signals from progressively higher chromospheric heights as a function of line depth. This provides confidence in the capability of bi-sector Doppler velocity analysis to resolve wave signatures in the chromosphere without the strict need of multi-line observations, and allows for an investigation of the development of the coherent pore oscillations as they move into the chromosphere.

\subsubsection{Inter-pore Coherency}
\label{coh}
Initial consideration was focused on whether the oscillations within each pore retained their photospheric coherence, i.e., remained in-phase with one another as they propagated into the chromosphere. Given the suppression of lower $p$-mode frequencies in the chromosphere, Fourier analysis was employed in the study of coherent wave signatures in order to provide better resulting frequency resolution, without the convolution effects of wavelet analyses. For this analysis and comparison between multiple time series at shared coherent frequencies, a 95\% confidence threshold was set as a baseline requirement for positive frequency detections.  

%%%%%%%%%%%%%%%%%%%%%%%%%%%%%%%%%%%%%%%%%%%%%%%
% FIGURE 7
%%%%%%%%%%%%%%%%%%%%%%%%%%%%%%%%%%%%%%%%%%%%%%%
\begin{figure}[t!]
\centering
  \includegraphics[trim=5mm 3mm 0mm 0mm, clip, width=0.85\columnwidth, angle=0]{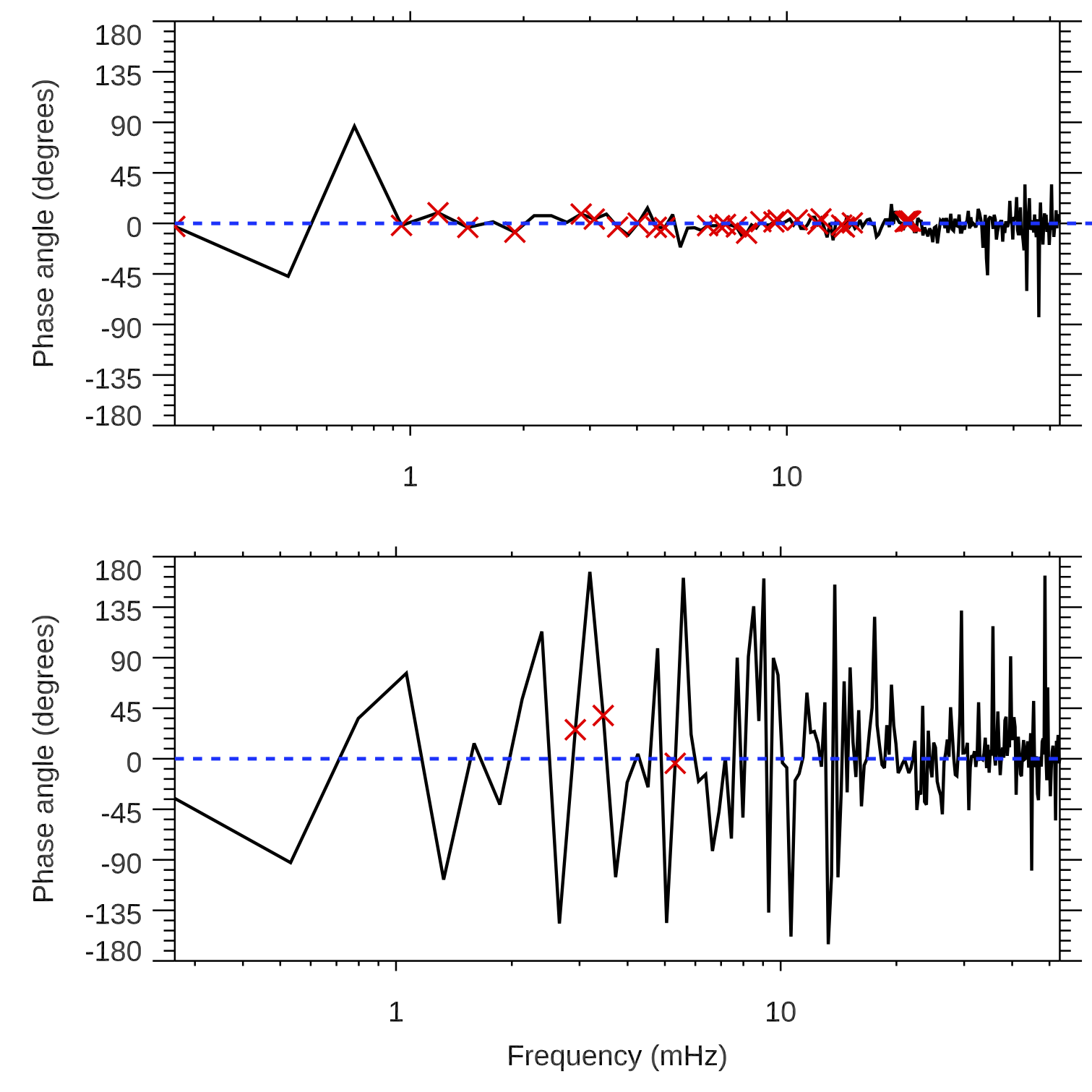}
\caption{Both panels depict the phase differences for the average (Eulerian) intensities between pores P3 \& P4 in different regions of the optical spectrum. The top panel highlights phases associated with the photospheric $4170${\,}{\AA} continuum channel, while the bottom panel documents the phases associated with the chromospheric line core of \CaIR spectral line. The red crosses indicate coherent frequencies above the 95\% confidence threshold, with a blue dashed line in each panel showing the location of zero degrees, indicative of `in-phase' oscillations.}
\label{rosa_ibis_phase}
\end{figure}
%%%%%%%%%%%%%%%%%%%%%%%%%%%%%%%%%%%%%%%%%%%%%%%
%%%%%%%%%%%%%%%%%%%%%%%%%%%%%%%%%%%%%%%%%%%%%%%
%%%%%%%%%%%%%%%%%%%%%%%%%%%%%%%%%%%%%%%%%%%%%%%

The \CaIR line core region was investigated since it probes the highest average chromospheric heights and provides the greatest contrast from the region studied in Section~\ref{photo}. To achieve this, bespoke pore perimeters were established from 8542.03{\,}{\AA} images, with selected contouring thresholds of P2 = $I_{\mathrm{mean}} - 1.60\sigma$, P3 = $I_{\mathrm{mean}} - 2.05\sigma$, P4 = $I_{\mathrm{mean}} - 2.20\sigma$, and P5 = $I_{\mathrm{mean}} - 1.50\sigma$ (see the lower panel of Figure~\ref{IBIS_fov}). 

From inspection, pores P3 \& P4 continue to be the largest and most complementary to each other, so a focus is placed on them. Figure~\ref{rosa_ibis_phase} displays the phase relationship between the Eulerian \CaIR line-core intensities of pores P3 \& P4 (lower panel), alongside a direct comparison to the ROSA 4170{\,}{\AA} continuum Eulerian intensity phases of the same pores (upper panel), as established in Section~\ref{photo}. The validity of the Fourier analysis technique is confirmed by the derived photospheric phases (upper panel of Figure~{\ref{rosa_ibis_phase}}), showing the same range of in-phase behaviour for frequencies $\ge1$~mHz as found in the previous wavelet analyses. Most notably, however, is the large discrepancy between the behaviours of the photospheric and chromospheric phase relationships, which can be seen by comparing the upper and lower panels of Figure~{\ref{rosa_ibis_phase}}. 

Of the frequencies displaying wave activity exceeding a 95\% confidence threshold in the \CaIR line core (red crosses in the lower panel of Figure~{\ref{rosa_ibis_phase}}), only the predominant 5~mHz signal exhibits any in-phase signatures between pores P3 \& P4, with the other identified frequencies falling $30-45\degr$ out-of-phase. Alongside this, the clearest contrast is found in the reduction in correlated chromospheric frequencies, with only three frequencies in the $3-5$~mHz range, which do not necessarily represent the strongest detected power spectral densities (see, e.g., Figure~{\ref{bisector_density}}). This behaviour is seen for the bi-sector Doppler velocities derived across all percentage line depths, with the sharp reduction in inter-pore in-phase signatures occurring around the cut-off height ($\sim$40\% line depth). The extinction of lower frequencies is to be expected, as a result of the cut-off effect within this region. However, no simple process can account for the chromospheric frequencies falling out-of-phase with one another between neighbouring pores, including powerful frequencies around 4~mHz, and those close to 7~mHz (see Figure~{\ref{bisector_density}}). The reduction in in-phase behaviour cannot be a result of instrumentation, since the $9.8$~s cadence of the chromospheric IBIS observations places the Nyquist frequency well above the frequency interval being discussed. Instead, the effect may be a result of pore structuring and the evolution experienced as the wave guides extend through the stratified atmosphere.

%%%%%%%%%%%%%%%%%%%%%%%%%%%%%%%%%%%%%%%%%%%%%%%
% FIGURE 8
%%%%%%%%%%%%%%%%%%%%%%%%%%%%%%%%%%%%%%%%%%%%%%%
\begin{figure}[t!]
\centering
  \includegraphics[trim=3mm 5mm 10mm 13mm, clip, width=0.8\columnwidth, angle=0]{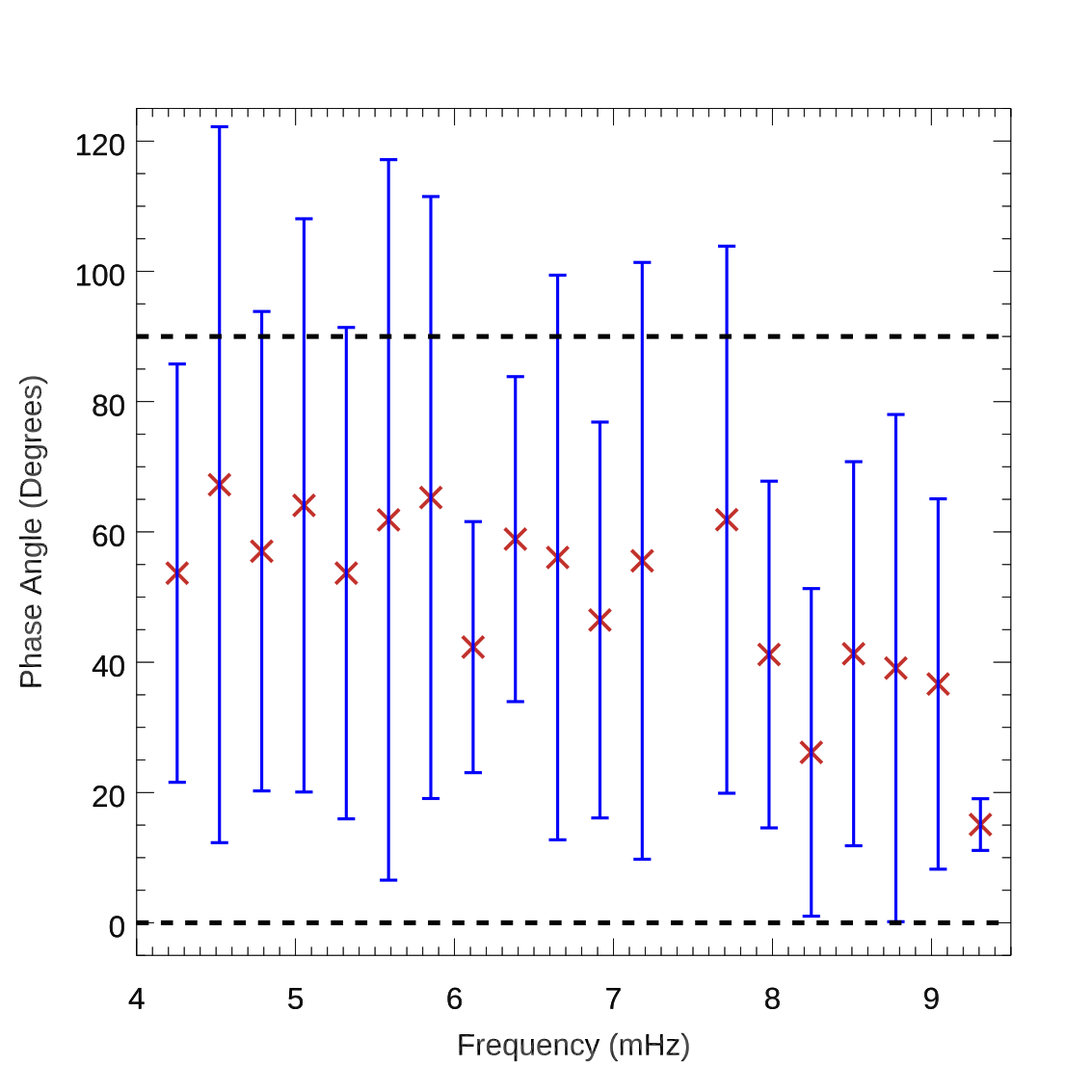}
\caption{The average phase differences (red crosses) between oscillations in the \CaIR line-of-sight velocity, {\vlos}, and associated line-core intensities of the pixels contained within pore P3. The error bars (blue) represent the standard deviation of the measured $I-V$ phases at each frequency, with horizontal dashed black lines highlighting $0{\degr}$ and $90{\degr}$ phase angles for reference.}
\label{iv_phase}
\end{figure}
%%%%%%%%%%%%%%%%%%%%%%%%%%%%%%%%%%%%%%%%%%%%%%%
%%%%%%%%%%%%%%%%%%%%%%%%%%%%%%%%%%%%%%%%%%%%%%%
%%%%%%%%%%%%%%%%%%%%%%%%%%%%%%%%%%%%%%%%%%%%%%%

Next, the phase relationships between the \CaIR line-core intensities, $I$, and the derived line-core Doppler velocities, {\vlos}, were studied for each pixel contained within the pores. The aggregated result for the $I-V$ phases within each pixel of pore P3 is displayed in Figure~\ref{iv_phase}, which is representative of the general trend across all chromospheric pore measurements captured in our observing sequence. From inspection of Figure~{\ref{iv_phase}}, the $I-V$ measurements are consistently out-of-phase, with average phase lags of $\sim40-70\degr$ across all frequencies. Furthermore, each frequency displays a significant spread of phases across the body of the pore, with standard deviations on the order of the mean phase value. Natural variance in $I-V$ is expected at chromospheric heights due to aspects of radiative damping \citep{Severino2013}, but not to the extent observed in Figure~\ref{iv_phase} indicative of a non-unified structuring of wave activity. In larger flux tubes, such as sunspots, certain wave modes have been observed to oscillate in a monolithic fashion radially across the tube, akin to a drum skin \citep{2012ApJ...757..160J, 2021A&A...649A.169S, Stan2022}. However, recent work has revealed that on small-scales, sunspots show a `corrugated' magnetic structure, with a range of fibrils permeating across the flux tube \citep{Rouppe2013, Yurch2014, Nelson2017}. It was shown that these fields cause phase shifts in non-linear magneto-acoustic shocks in the umbra \citep{Henriques2020}. In the case of the pores presented in the current study, no sub-regions exhibit spatial coherence, suggesting that the pores manifest as a collection of fractured wave guides in the chromosphere, rather than as a single monolithic tube as in the photosphere. This is consistent with the corrugated model of \citet{Henriques2020}, where the corrugation leads to the annihilation of in-phase inter-pore frequencies, as the oscillations within each pore propagate in their own unique manner that is governed by magnetic field inclination changes across the pore and the small-scale plasma flows associated with a corrugated atmosphere. Hence, despite the wave power existing across all flux tubes, they no longer share the commonality they possessed upon emergence at photospheric heights.

%%%%%%%%%%%%%%%%%%%%%%%%%%%%%%%%%%%%%%%%%%%%%%%
% FIGURE 9
%%%%%%%%%%%%%%%%%%%%%%%%%%%%%%%%%%%%%%%%%%%%%%%
\begin{figure}[t!]
\centering
  \includegraphics[trim=22mm 5mm 5mm 10mm, clip, width=0.8\columnwidth, angle=0]{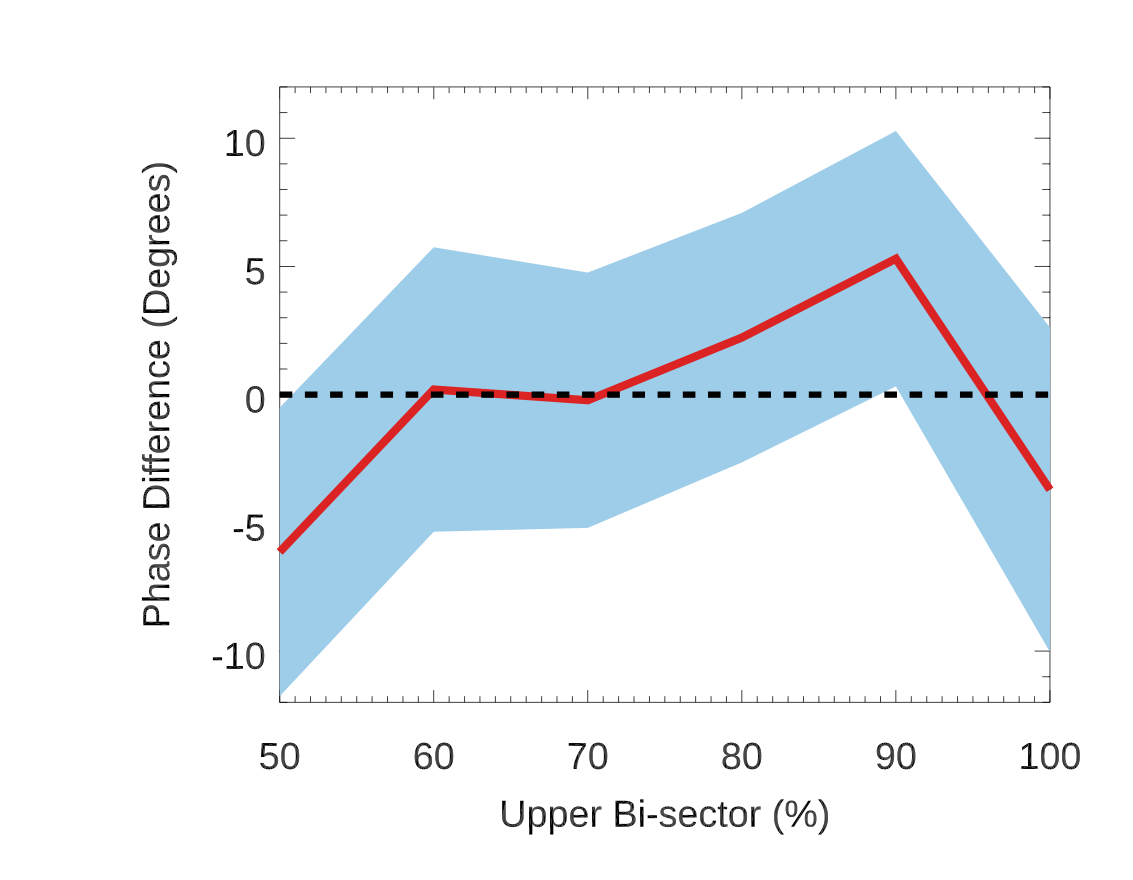}
\caption{The phase differences for frequencies detected in the displayed bi-sector and the 10\% lower line depth, where a negative phase angle indicates the wave is detected at the lower line depth first. The average phase difference is plotted in red, with the standard deviation of each height differential shown in blue.}
\label{ibis_phase}
\end{figure}
%%%%%%%%%%%%%%%%%%%%%%%%%%%%%%%%%%%%%%%%%%%%%%%
%%%%%%%%%%%%%%%%%%%%%%%%%%%%%%%%%%%%%%%%%%%%%%%
%%%%%%%%%%%%%%%%%%%%%%%%%%%%%%%%%%%%%%%%%%%%%%%

\subsubsection{Wave Propagation}
\citetalias{GM21} reported notable energy damping of oscillations in photospheric {\vlos} signatures in all pores as they passed through the lower atmosphere, up to an atmospheric height of approximately 500~km. The elimination of inter-pore coherency discussed in Section~\ref{coh} doesn't preclude the propagation of the detected wave power within each pore, so it is pertinent to study the continuing viability of these waves as energy conduits. 

The phase lags between line-core \CaIR intensity and its associated line-of-sight Doppler velocity, {\vlos}, as discussed in Section~\ref{coh}, has previously been used to determine the precise mode of compressible sausage waves \citep[e.g.,][]{Tsap2016}. Through the derived relationships presented by \citet{Moreels2013a}, the slow, propagating sausage mode observed by \citetalias{GM21} would produce an in-phase $I-V$ relationship, i.e., a $0\degr$ phase lag, which from inspection of Figure~\ref{iv_phase} is not necessarily the case at chromospheric heights. \citet{Moreels2013a} allows for phase solutions in the range of $0 - 90\degr$, however, the lack of consensus on a single phase angle for each frequency (see Figure~{\ref{iv_phase}}) suggests that the exact mode of propagation in the chromosphere is not discernible through $I-V$ phase alone. It must be noted that \citet{Moreels2013a} based their model on a photospheric flux tube, therefore, it is not unexpected that the derived relationships for compressible modes do not hold in the more complex chromosphere. Instead, bi-sector Doppler velocities derived from consecutively increasing percentage line depths must be employed directly to search for signatures of propagation. 

To directly assess the propagation of wave signatures in the lower chromosphere of the pores, the Eulerian bi-sector velocities for each pore at neighbouring percentage line depths were analysed. Due to the persistence and size of pores P3 \& P4, we limited our study to these long-lived and stable magnetic features. Furthermore, for consistency to previous analyses, we only select those frequencies which exceed the 95\% confidence threshold across all percentage line depth comparisons, i.e., $40-50$\%, $50-60$\%, $60-70$\%, $70-80$\%, $80-90$\%, and $90-100$\% (where $100$\% line depth corresponds to the \CaIR line core). In total, 11 frequencies between $4-6$~mHz were identifed in pores P3 \& P4.

For the detection of definitely upwardly propagating waves, the derived phase angles would need to display clear negative phase angles (i.e., the 50\% line depth wave signals should trail their 40\% line depth counterparts). As can be seen in Figure~{\ref{ibis_phase}}, the measured phase angles between neighbouring \CaIR percentage line depths do not show clear negative phase lags, with the mean values at each line depth interval distributed around zero degrees. Indeed, including the standard errors, each line depth interval straddles the zero degree phase threshold, suggesting the presence of standing mode waves in the pores at chromospheric heights. Standing compressible modes have previously been observed in chromospheric magnetic flux tubes, with \citet{Freij2016} postulating that the reflection occurs at the transition region boundary, and could be indicative of a chromospheric resonator \citep{Jess2020, 2021NatAs...5....5J, 2020ApJ...900L..29F, 2021NatAs...5....2F}. The viability of this reflection region can be assessed through calculating the typical wavelength of these standing modes. The average tube speed of the pores in the high photosphere, as inferred by \citetalias{GM21}, of $\sim 8$~km/s can be combined with the predominant $5$~mHz frequency through $\lambda = {v_{T}} / f$ to give an approximate wavelength of $1600$~km. This places a further node at $\sim 2400$~km, and considering the increase in tube speed associated with the chromosphere leading to an increased wavelength, there is ample opportunity for a standing mode reflection to form at transition region heights.  

\section{Conclusion}

The nature of wave generation and chromospheric propagation within a unique cluster of five magnetic pores is presented by employing wavelet and Fourier analysis to a novel configuration of high-resolution, multi-wavelength observations. This study has revealed much about the characteristics of pores as wave guides, alongside providing new mechanisms for multi-height chromospheric diagnostics. In particular,

\begin{itemize}
    \item Coherent oscillations in the intensity emission and cross-sectional area, consistent with slow sausage modes, were detected across all pores in the photosphere, with robust analysis concluding that instrumental noise or any mitigation techniques applied do not facilitate such coherence. The remarkable coherence across the pores is indicative of a common driver, acting on the flux tubes below the solar surface where it is believed they constitute a monolithic flux rope.  
    \item \CaIR bi-sector analysis was employed to extract Doppler velocity signals across a range of percentage line depths and to test their viability to probe discrete heights in the lower solar atmosphere. It was found that the bi-sector velocities probe an atmospheric height above the cut-off region that prohibits the propagation of 3~mHz signals, and that each increasing percentage line depth exhibits a clear wave power transition towards dominant 5~mHz power, which has only been seen with such fidelity in simulations \citep[e.g.,][]{Felipe2019}. The increasing 5~mHz wave power found as a function of percentage line depth is indicative of the bi-sector velocities sampling progressively higher atmospheric heights, which was confirmed through response function calculations of the \CaIR line. 
    \item The extinction of the ample photospheric wave coherency across a range of frequencies in the chromosphere was seen in 
    \CaIR line-core intensities and bi-sector Doppler velocities, with only the 5~mHz $p$-mode peak retaining any semblance of correlation. Phase relationships between intensity and Doppler velocity revealed the fractured nature of intra-pore oscillations, with a lack of uniformity in wave phases indicative of unique propagation parameters across the surface of a single pore structure, consistent with the first detection of a corrugated atmosphere in the chromosphere of a magnetic pore.
    \item The waves displayed propagation characteristics consistent with standing modes through Fourier analysis of bi-sector velocities derived from adjacent \CaIR percentage line depths spanning from the photospheric wings of the profile to the chromospheric line-core. Eleven frequencies, each displaying $\ge95$\% confidence levels, spanned all percentage line depths in pores P3 \& P4, with phase angles between neighbouring percentage line depths around zero degrees across all frequencies, indicative of standing modes formed due to a reflection layer at the base of the transition region.
\end{itemize}

The distinctive arrangement of this active region has allowed for an unprecedented study into the effects of pore structuring on wave transportation into the chromosphere, given the injection of equivalent, coherent waves at the solar surface. We have shown that the fine-scale differences in each magnetic flux tube can cause notable variations in the evolution of waves contained within. The phase differences seen in Figure~\ref{ibis_phase} indicate the corrugation of the chromospheric pore atmosphere, consistent with the observations and models of \citep{Henriques2020}. In their case, the larger density gradients of sunspots produced macroscopic shock events capable of highlighting the corrugation due to horizontal fibrils. As discussed in Section~\ref{sec:bi}, there are no observable shock events within the pores, however this does not prohibit the existence of the small-scale brightenings associated with umbral corrugation since the filling factor of their associated pixel may be dominated by quiescent plasma. Indeed, \citet{Rouppe2013} confirmed the fibrillar nature of umbrae using smaller sunspots with partial penumbra, indicating that the effect can be scaled down to smaller structures. Rather than relying on macroscopic shocks, the lack of phase across each pore provides the first evidence of corrugation in small-scale magnetic structures. Magnetic field corrugation across a range of magnetic structures would have implications on the energy transport and deposition through wave motion. However, semi-empirical modeling through inversions, similar to the robust methods of \citet{Henriques2020}, would be advantageous in future to fully assess the impact of corrugation on magnetic pores. This behavior could also be probed by considering the pores as coupled systems, with a photospheric input and chromospheric output. The lack of a linear change in phase as a function of height implies that the pores form non-linear systems, which can be identified and characterised with models such as NARMAX \citep{Chen1989}. This would provide an unprecedented view of magnetic flux tubes as systems, and provide explanations for the complex phase relationships observed in this data. However, equivalent absolute measurements would be needed of the input and outputs of the system, such as LOS velocity, which is not possible for this study. Future observing runs that incorporate spectral imaging throughout the lower solar atmosphere could provide a novel investigation of pores as non-linear systems.

Propagating waves are often thought necessary for the rapid damping observed by \citetalias{GM21}. However, \citetalias{GM21} examined $\sim3$~mHz photospheric signals, which do not readily traverse the chromospheric boundary due to the atmospheric cut-off. Despite this, damping can still occur in standing modes through partial transmission of wave power at the upper reflective boundary, or geometric spreading and lateral wave leakage, as put forward by \citet{Riedl2021}. Therefore, the confirmation of standing modes in chromospheric flux tubes does not preclude energy transport and damping, and as such active region configurations such as these can have an influence on energy dissipation and chromospheric heating.

Looking forward, bi-sector analysis using the \CaIR line has been shown to be a viable tool for extracting multi-height chromospheric information from a single observable in the case of these pores, though care must be taken to account for density enhancements in the line wings and/or non-linear effects. 
%Their viability for use with small magnetic flux tubes has been tentatively shown, provided localised density enhancements, and non-linearities due to shocks, light bridges or plage causing line core emission are discounted.
Certainly for sunspots, bi-sector velocity studies would be better suited for spectral lines with less intensity responsiveness, such as H$\alpha$. For wave observers using cutting-edge suites such as DKIST, bi-sectors may provide an avenue for observing dynamic signals in the atmosphere without sacrificing temporal or spatial instrumental resolution. These modern observatories can also provide further insight into the variation in pore wave conduit behavior. Allied to this, modeling of multiple three-dimensional flux tubes in close proximity is in its infancy \citep[e.g.,][]{Snow2018}. However, there is scope for simulating scenarios similar to this active region to further constrain how flux tubes can differ, and whether they are a direct influence on one another across differing atmospheric heights. Allied to this, the future direction of observing campaigns can focus on more complex regions of monopolar flux tubes, to assess how unique the configuration and wave activity observed in this study truly is. 

\begin{acknowledgements}
\noindent 
The authors acknowledge H. Schunker for valuable discussions.
% Sam & Dave
S.D.T.G. and D.B.J. are grateful to Invest NI and Randox Laboratories Ltd. for the award of a Research \& Development Grant (059RDEN-1), in addition to the UK STFC for the consolidated grant ST/T00021X/1. 
% Marco
% Shahin
SJ acknowledges support from the European Research Council under the European Union Horizon 2020 research and innovation program (grant agreement No. 682462) and from the Research Council of Norway through its Centres of Excellence scheme (project No. 262622). 
% Peter
% Paul
% Conor
CDM would like to thank the Northern Ireland Department for the Economy for the award of a PhD studentship. 
% Caitlin
CAG-M is grateful to Randox Laboratories Ltd. for the award of a PhD studentship. 
The Dunn Solar Telescope at Sacramento Peak/NM was operated by the National Solar Observatory (NSO). NSO is operated by the Association of Universities for Research in Astronomy (AURA), Inc., under cooperative agreement with the National Science Foundation (NSF). 
The authors wish to acknowledge scientific discussions with the Waves in the Lower Solar Atmosphere (WaLSA; \href{www.WaLSA.team}{www.WaLSA.team}) team, which is supported by the Research Council of Norway (project number 262622), and The Royal Society through the award of funding to host the Theo Murphy Discussion Meeting ``High-resolution wave dynamics in the lower solar atmosphere'' (grant Hooke18b/SCTM). 
\end{acknowledgements}

\bibliographystyle{aasjournal.bst}
\bibliography{references}

\begin{thebibliography}{}
\expandafter\ifx\csname natexlab\endcsname\relax\def\natexlab#1{#1}\fi
\providecommand{\url}[1]{\href{#1}{#1}}
\providecommand{\dodoi}[1]{doi:~\href{http://doi.org/#1}{\nolinkurl{#1}}}
\providecommand{\doeprint}[1]{\href{http://ascl.net/#1}{\nolinkurl{http://ascl.net/#1}}}
\providecommand{\doarXiv}[1]{\href{https://arxiv.org/abs/#1}{\nolinkurl{https://arxiv.org/abs/#1}}}

\bibitem[{{Albidah} {et~al.}(2021){Albidah}, {Brevis}, {Fedun}, {Ballai},
  {Jess}, {Stangalini}, {Higham}, \& {Verth}}]{2021RSPTA.37900181A}
{Albidah}, A.~B., {Brevis}, W., {Fedun}, V., {et~al.} 2021, Philosophical
  Transactions of the Royal Society of London Series A, 379, 20200181,
  \dodoi{10.1098/rsta.2020.0181}

\bibitem[{{Banerjee} {et~al.}(2007){Banerjee}, {Erd{\'e}lyi}, {Oliver}, \&
  {O'Shea}}]{Ban2007}
{Banerjee}, D., {Erd{\'e}lyi}, R., {Oliver}, R., \& {O'Shea}, E. 2007,
  \solphys, 246, 3, \dodoi{10.1007/s11207-007-9029-z}

\bibitem[{{Beck} \& {Choudhary}(2020)}]{Beck2020}
{Beck}, C., \& {Choudhary}, D.~P. 2020, \apj, 891, 119,
  \dodoi{10.3847/1538-4357/ab75bd}

\bibitem[{{Bel} \& {Leroy}(1977)}]{Bel1977}
{Bel}, N., \& {Leroy}, B. 1977, \aap, 55, 239

\bibitem[{{Braun} {et~al.}(1988){Braun}, {Duvall}, \& {Labonte}}]{Braun1988}
{Braun}, D.~C., {Duvall}, T.~L., J., \& {Labonte}, B.~J. 1988, \apj, 335, 1015,
  \dodoi{10.1086/166988}

\bibitem[{{Carlsson} \& {Stein}(1997)}]{Carlsson1997}
{Carlsson}, M., \& {Stein}, R.~F. 1997, \apj, 481, 500, \dodoi{10.1086/304043}

\bibitem[{{Cauzzi} {et~al.}(2008){Cauzzi}, {Reardon}, {Uitenbroek},
  {Cavallini}, {Falchi}, {Falciani}, {Janssen}, {Rimmele}, {Vecchio}, \&
  {W{\"o}ger}}]{Cauzzi2008}
{Cauzzi}, G., {Reardon}, K.~P., {Uitenbroek}, H., {et~al.} 2008, \aap, 480,
  515, \dodoi{10.1051/0004-6361:20078642}

\bibitem[{{Cavallini}(2006)}]{Cav2006}
{Cavallini}, F. 2006, \solphys, 236, 415, \dodoi{10.1007/s11207-006-0103-8}

\bibitem[{{Centeno} {et~al.}(2006){Centeno}, {Collados}, \& {Trujillo
  Bueno}}]{Centeno2006}
{Centeno}, R., {Collados}, M., \& {Trujillo Bueno}, J. 2006, \apj, 640, 1153,
  \dodoi{10.1086/500185}

\bibitem[{{Centeno} {et~al.}(2009){Centeno}, {Collados}, \& {Trujillo
  Bueno}}]{Centeno2009}
---. 2009, \apj, 692, 1211, \dodoi{10.1088/0004-637X/692/2/1211}

\bibitem[{{Chae} {et~al.}(2013){Chae}, {Park}, {Ahn}, {Yang}, {Park}, {Cho}, \&
  {Cao}}]{Chae2013}
{Chae}, J., {Park}, H.-M., {Ahn}, K., {et~al.} 2013, \solphys, 288, 89,
  \dodoi{10.1007/s11207-013-0313-9}

\bibitem[{{Chen} {et~al.}(2017){Chen}, {Rempel}, \& {Fan}}]{Feng2017}
{Chen}, F., {Rempel}, M., \& {Fan}, Y. 2017, \apj, 846, 149,
  \dodoi{10.3847/1538-4357/aa85a0}

\bibitem[{Chen \& Billings(1989)}]{Chen1989}
Chen, S., \& Billings, S.~A. 1989, International Journal of Control, 49, 1013,
  \dodoi{10.1080/00207178908559683}

\bibitem[{{Cheung} {et~al.}(2010){Cheung}, {Rempel}, {Title}, \&
  {Sch{\"u}ssler}}]{Cheung2010}
{Cheung}, M.~C.~M., {Rempel}, M., {Title}, A.~M., \& {Sch{\"u}ssler}, M. 2010,
  \apj, 720, 233, \dodoi{10.1088/0004-637X/720/1/233}

\bibitem[{{Cho} {et~al.}(2013){Cho}, {Bong}, {Chae}, {Kim}, {Park}, \&
  {Katsukawa}}]{Cho2013}
{Cho}, K.~S., {Bong}, S.~C., {Chae}, J., {et~al.} 2013, \solphys, 288, 23,
  \dodoi{10.1007/s11207-012-0196-1}

\bibitem[{{Cho} {et~al.}(2015){Cho}, {Bong}, {Nakariakov}, {Lim}, {Park},
  {Chae}, {Yang}, {Park}, \& {Yurchyshyn}}]{Cho2015}
{Cho}, K.~S., {Bong}, S.~C., {Nakariakov}, V.~M., {et~al.} 2015, \apj, 802, 45,
  \dodoi{10.1088/0004-637X/802/1/45}

\bibitem[{{De Moortel} \& {Browning}(2015)}]{Moortel2015}
{De Moortel}, I., \& {Browning}, P. 2015, Philosophical Transactions of the
  Royal Society of London Series A, 373, 20140269,
  \dodoi{10.1098/rsta.2014.0269}

\bibitem[{{Dorotovi{\v{c}}} {et~al.}(2008){Dorotovi{\v{c}}}, {Erd{\'e}lyi}, \&
  {Karlovsk{\'y}}}]{Doro2008}
{Dorotovi{\v{c}}}, I., {Erd{\'e}lyi}, R., \& {Karlovsk{\'y}}, V. 2008, in Waves
  \& Oscillations in the Solar Atmosphere: Heating and Magneto-Seismology, ed.
  R.~{Erd{\'e}lyi} \& C.~A. {Mendoza-Briceno}, Vol. 247, 351--354,
  \dodoi{10.1017/S174392130801507X}

\bibitem[{{Edwin} \& {Roberts}(1983)}]{Edwin1983}
{Edwin}, P.~M., \& {Roberts}, B. 1983, \solphys, 88, 179,
  \dodoi{10.1007/BF00196186}

\bibitem[{{Felipe}(2019)}]{Felipe2019}
{Felipe}, T. 2019, \aap, 627, A169, \dodoi{10.1051/0004-6361/201935784}

\bibitem[{{Felipe}(2021)}]{2021NatAs...5....2F}
---. 2021, Nature Astronomy, 5, 2, \dodoi{10.1038/s41550-020-1157-5}

\bibitem[{{Felipe} {et~al.}(2020){Felipe}, {Kuckein}, {Gonz{\'a}lez Manrique},
  {Milic}, \& {Sangeetha}}]{2020ApJ...900L..29F}
{Felipe}, T., {Kuckein}, C., {Gonz{\'a}lez Manrique}, S.~J., {Milic}, I., \&
  {Sangeetha}, C.~R. 2020, \apjl, 900, L29, \dodoi{10.3847/2041-8213/abb1a5}

\bibitem[{{Felipe} {et~al.}(2018){Felipe}, {Kuckein}, \& {Thaler}}]{Felipe2018}
{Felipe}, T., {Kuckein}, C., \& {Thaler}, I. 2018, \aap, 617, A39,
  \dodoi{10.1051/0004-6361/201833155}

\bibitem[{{Felipe} \& {Sangeetha}(2020)}]{Felipe2020}
{Felipe}, T., \& {Sangeetha}, C.~R. 2020, \aap, 640, A4,
  \dodoi{10.1051/0004-6361/202038387}

\bibitem[{{Fossum} \& {Carlsson}(2005{\natexlab{a}})}]{Fossum2005}
{Fossum}, A., \& {Carlsson}, M. 2005{\natexlab{a}}, \nat, 435, 919,
  \dodoi{10.1038/nature03695}

\bibitem[{{Fossum} \& {Carlsson}(2005{\natexlab{b}})}]{2005ApJ...625..556F}
---. 2005{\natexlab{b}}, \apj, 625, 556, \dodoi{10.1086/429614}

\bibitem[{{Freij} {et~al.}(2016){Freij}, {Dorotovi{\v{c}}}, {Morton},
  {Ruderman}, {Karlovsk{\'y}}, \& {Erd{\'e}lyi}}]{Freij2016}
{Freij}, N., {Dorotovi{\v{c}}}, I., {Morton}, R.~J., {et~al.} 2016, \apj, 817,
  44, \dodoi{10.3847/0004-637X/817/1/44}

\bibitem[{{Garcia de La Rosa}(1987)}]{Garcia1987}
{Garcia de La Rosa}, J.~I. 1987, \solphys, 112, 49, \dodoi{10.1007/BF00148486}

\bibitem[{{Gilchrist-Millar} {et~al.}(2021){Gilchrist-Millar}, {Jess}, {Grant},
  {Keys}, {Beck}, {Jafarzadeh}, {Riedl}, {Van Doorsselaere}, \& {Ruiz
  Cobo}}]{GM21}
{Gilchrist-Millar}, C.~A., {Jess}, D.~B., {Grant}, S. D.~T., {et~al.} 2021,
  Philosophical Transactions of the Royal Society of London Series A, 379,
  20200172, \dodoi{10.1098/rsta.2020.0172}

\bibitem[{{Giovanelli} {et~al.}(1978){Giovanelli}, {Harvey}, \&
  {Livingston}}]{Gio1978}
{Giovanelli}, R.~G., {Harvey}, J.~W., \& {Livingston}, W.~C. 1978, \solphys,
  58, 347, \dodoi{10.1007/BF00157281}

\bibitem[{{Gonz{\'a}lez Manrique} {et~al.}(2020){Gonz{\'a}lez Manrique},
  {Quintero Noda}, {Kuckein}, {Ruiz Cobo}, \& {Carlsson}}]{2020A&A...634A..19G}
{Gonz{\'a}lez Manrique}, S.~J., {Quintero Noda}, C., {Kuckein}, C., {Ruiz
  Cobo}, B., \& {Carlsson}, M. 2020, \aap, 634, A19,
  \dodoi{10.1051/0004-6361/201937274}

\bibitem[{{Grant} {et~al.}(2015){Grant}, {Jess}, {Moreels}, {Morton},
  {Christian}, {Giagkiozis}, {Verth}, {Fedun}, {Keys}, {Van Doorsselaere}, \&
  {Erd{\'e}lyi}}]{Grant2015}
{Grant}, S.~D.~T., {Jess}, D.~B., {Moreels}, M.~G., {et~al.} 2015, \apj, 806,
  132, \dodoi{10.1088/0004-637X/806/1/132}

\bibitem[{{Grant} {et~al.}(2018){Grant}, {Jess}, {Zaqarashvili}, {Beck},
  {Socas-Navarro}, {Aschwanden}, {Keys}, {Christian}, {Houston}, \&
  {Hewitt}}]{Grant2018}
{Grant}, S. D.~T., {Jess}, D.~B., {Zaqarashvili}, T.~V., {et~al.} 2018, Nature
  Physics, 14, 480, \dodoi{10.1038/s41567-018-0058-3}

\bibitem[{{Henriques} {et~al.}(2017){Henriques}, {Mathioudakis},
  {Socas-Navarro}, \& {de la Cruz Rodr{\'\i}guez}}]{Henriques2017}
{Henriques}, V.~M.~J., {Mathioudakis}, M., {Socas-Navarro}, H., \& {de la Cruz
  Rodr{\'\i}guez}, J. 2017, \apj, 845, 102, \dodoi{10.3847/1538-4357/aa7ca4}

\bibitem[{{Henriques} {et~al.}(2020){Henriques}, {Nelson}, {Rouppe van der
  Voort}, \& {Mathioudakis}}]{Henriques2020}
{Henriques}, V. M.~J., {Nelson}, C.~J., {Rouppe van der Voort}, L. H.~M., \&
  {Mathioudakis}, M. 2020, \aap, 642, A215, \dodoi{10.1051/0004-6361/202038538}

\bibitem[{{Jacoutot} {et~al.}(2008){Jacoutot}, {Kosovichev}, {Wray}, \&
  {Mansour}}]{Jacoutot2008}
{Jacoutot}, L., {Kosovichev}, A.~G., {Wray}, A., \& {Mansour}, N.~N. 2008,
  \apjl, 684, L51, \dodoi{10.1086/592042}

\bibitem[{{Jess} {et~al.}(2007){Jess}, {Andi{\'c}}, {Mathioudakis},
  {Bloomfield}, \& {Keenan}}]{Jess2007}
{Jess}, D.~B., {Andi{\'c}}, A., {Mathioudakis}, M., {Bloomfield}, D.~S., \&
  {Keenan}, F.~P. 2007, \aap, 473, 943, \dodoi{10.1051/0004-6361:20077142}

\bibitem[{{Jess} {et~al.}(2012{\natexlab{a}}){Jess}, {De Moortel},
  {Mathioudakis}, {Christian}, {Reardon}, {Keys}, \&
  {Keenan}}]{2012ApJ...757..160J}
{Jess}, D.~B., {De Moortel}, I., {Mathioudakis}, M., {et~al.}
  2012{\natexlab{a}}, \apj, 757, 160, \dodoi{10.1088/0004-637X/757/2/160}

\bibitem[{{Jess} {et~al.}(2010{\natexlab{a}}){Jess}, {Mathioudakis},
  {Christian}, {Crockett}, \& {Keenan}}]{2010ApJ...719L.134J}
{Jess}, D.~B., {Mathioudakis}, M., {Christian}, D.~J., {Crockett}, P.~J., \&
  {Keenan}, F.~P. 2010{\natexlab{a}}, \apjl, 719, L134,
  \dodoi{10.1088/2041-8205/719/2/L134}

\bibitem[{{Jess} {et~al.}(2010{\natexlab{b}}){Jess}, {Mathioudakis},
  {Christian}, {Keenan}, {Ryans}, \& {Crockett}}]{Jess2010}
{Jess}, D.~B., {Mathioudakis}, M., {Christian}, D.~J., {et~al.}
  2010{\natexlab{b}}, \solphys, 261, 363, \dodoi{10.1007/s11207-009-9500-0}

\bibitem[{{Jess} {et~al.}(2014){Jess}, {Mathioudakis}, \& {Keys}}]{Jess2014}
{Jess}, D.~B., {Mathioudakis}, M., \& {Keys}, P.~H. 2014, \apj, 795, 172,
  \dodoi{10.1088/0004-637X/795/2/172}

\bibitem[{{Jess} {et~al.}(2015){Jess}, {Morton}, {Verth}, {Fedun}, {Grant}, \&
  {Giagkiozis}}]{Jess2015}
{Jess}, D.~B., {Morton}, R.~J., {Verth}, G., {et~al.} 2015, \ssr, 190, 103,
  \dodoi{10.1007/s11214-015-0141-3}

\bibitem[{{Jess} {et~al.}(2012{\natexlab{b}}){Jess}, {Shelyag}, {Mathioudakis},
  {Keys}, {Christian}, \& {Keenan}}]{Jess2012}
{Jess}, D.~B., {Shelyag}, S., {Mathioudakis}, M., {et~al.} 2012{\natexlab{b}},
  \apj, 746, 183, \dodoi{10.1088/0004-637X/746/2/183}

\bibitem[{{Jess} {et~al.}(2021){Jess}, {Snow}, {Fleck}, {Stangalini}, \&
  {Jafarzadeh}}]{2021NatAs...5....5J}
{Jess}, D.~B., {Snow}, B., {Fleck}, B., {Stangalini}, M., \& {Jafarzadeh}, S.
  2021, Nature Astronomy, 5, 5, \dodoi{10.1038/s41550-020-1158-4}

\bibitem[{{Jess} {et~al.}(2016){Jess}, {Reznikova}, {Ryans}, {Christian},
  {Keys}, {Mathioudakis}, {Mackay}, {Krishna Prasad}, {Banerjee}, {Grant},
  {Yau}, \& {Diamond}}]{2016NatPh..12..179J}
{Jess}, D.~B., {Reznikova}, V.~E., {Ryans}, R. S.~I., {et~al.} 2016, Nature
  Physics, 12, 179, \dodoi{10.1038/nphys3544}

\bibitem[{{Jess} {et~al.}(2017){Jess}, {Van Doorsselaere}, {Verth}, {Fedun},
  {Krishna Prasad}, {Erd{\'e}lyi}, {Keys}, {Grant}, {Uitenbroek}, \&
  {Christian}}]{2017ApJ...842...59J}
{Jess}, D.~B., {Van Doorsselaere}, T., {Verth}, G., {et~al.} 2017, \apj, 842,
  59, \dodoi{10.3847/1538-4357/aa73d6}

\bibitem[{{Jess} {et~al.}(2019){Jess}, {Dillon}, {Kirk}, {Reale},
  {Mathioudakis}, {Grant}, {Christian}, {Keys}, {Krishna Prasad}, \&
  {Houston}}]{Jess2019}
{Jess}, D.~B., {Dillon}, C.~J., {Kirk}, M.~S., {et~al.} 2019, \apj, 871, 133,
  \dodoi{10.3847/1538-4357/aaf8ae}

\bibitem[{{Jess} {et~al.}(2020){Jess}, {Snow}, {Houston}, {Botha}, {Fleck},
  {Krishna Prasad}, {Asensio Ramos}, {Morton}, {Keys}, {Jafarzadeh},
  {Stangalini}, {Grant}, \& {Christian}}]{Jess2020}
{Jess}, D.~B., {Snow}, B., {Houston}, S.~J., {et~al.} 2020, Nature Astronomy,
  4, 220, \dodoi{10.1038/s41550-019-0945-2}

\bibitem[{{Kato} {et~al.}(2011){Kato}, {Steiner}, {Steffen}, \&
  {Suematsu}}]{Kato2011}
{Kato}, Y., {Steiner}, O., {Steffen}, M., \& {Suematsu}, Y. 2011, \apjl, 730,
  L24, \dodoi{10.1088/2041-8205/730/2/L24}

\bibitem[{{Keys} {et~al.}(2014){Keys}, {Mathioudakis}, {Jess}, {Mackay}, \&
  {Keenan}}]{Keys2014}
{Keys}, P.~H., {Mathioudakis}, M., {Jess}, D.~B., {Mackay}, D.~H., \& {Keenan},
  F.~P. 2014, \aap, 566, A99, \dodoi{10.1051/0004-6361/201322987}

\bibitem[{{Keys} {et~al.}(2018){Keys}, {Morton}, {Jess}, {Verth}, {Grant},
  {Mathioudakis}, {Mackay}, {Doyle}, {Christian}, {Keenan}, \&
  {Erd{\'e}lyi}}]{Keys2018}
{Keys}, P.~H., {Morton}, R.~J., {Jess}, D.~B., {et~al.} 2018, \apj, 857, 28,
  \dodoi{10.3847/1538-4357/aab432}

\bibitem[{{Khomenko} \& {Collados}(2015)}]{Khom2015}
{Khomenko}, E., \& {Collados}, M. 2015, Living Reviews in Solar Physics, 12, 6,
  \dodoi{10.1007/lrsp-2015-6}

\bibitem[{{Kitiashvili} {et~al.}(2011){Kitiashvili}, {Kosovichev}, {Mansour},
  \& {Wray}}]{Kiti2011}
{Kitiashvili}, I.~N., {Kosovichev}, A.~G., {Mansour}, N.~N., \& {Wray}, A.~A.
  2011, \solphys, 268, 283, \dodoi{10.1007/s11207-010-9679-0}

\bibitem[{{Krishna Prasad} {et~al.}(2015){Krishna Prasad}, {Jess}, \&
  {Khomenko}}]{Krishna2015}
{Krishna Prasad}, S., {Jess}, D.~B., \& {Khomenko}, E. 2015, \apjl, 812, L15,
  \dodoi{10.1088/2041-8205/812/1/L15}

\bibitem[{{Kulander} \& {Jefferies}(1966)}]{Kulander1966}
{Kulander}, J.~L., \& {Jefferies}, J.~T. 1966, \apj, 146, 194,
  \dodoi{10.1086/148868}

\bibitem[{{Leenaarts} {et~al.}(2014){Leenaarts}, {de la Cruz Rodr{\'\i}guez},
  {Kochukhov}, \& {Carlsson}}]{Leen2014}
{Leenaarts}, J., {de la Cruz Rodr{\'\i}guez}, J., {Kochukhov}, O., \&
  {Carlsson}, M. 2014, \apjl, 784, L17, \dodoi{10.1088/2041-8205/784/1/L17}

\bibitem[{{Leighton} {et~al.}(1962){Leighton}, {Noyes}, \& {Simon}}]{Leigh1962}
{Leighton}, R.~B., {Noyes}, R.~W., \& {Simon}, G.~W. 1962, \apj, 135, 474,
  \dodoi{10.1086/147285}

\bibitem[{{Lites}(1984)}]{Lites1984}
{Lites}, B.~W. 1984, \apj, 277, 874, \dodoi{10.1086/161758}

\bibitem[{{Lites} {et~al.}(1982){Lites}, {White}, \& {Packman}}]{Lites1982}
{Lites}, B.~W., {White}, O.~R., \& {Packman}, D. 1982, \apj, 253, 386,
  \dodoi{10.1086/159642}

\bibitem[{{MacBride} \& {Jess}(2020)}]{Macbride2020}
{MacBride}, C.~D., \& {Jess}, D.~B. 2020, {MCALF v0.1}, v0.1,  Zenodo,
  \dodoi{10.5281/zenodo.3924527}

\bibitem[{MacBride \& Jess(2021)}]{MacBride2021b}
MacBride, C.~D., \& Jess, D.~B. 2021, Journal of Open Source Software, 6, 3265,
  \dodoi{10.21105/joss.03265}

\bibitem[{{MacBride} {et~al.}(2021){MacBride}, {Jess}, {Grant}, {Khomenko},
  {Keys}, \& {Stangalini}}]{Macbride2021a}
{MacBride}, C.~D., {Jess}, D.~B., {Grant}, S. D.~T., {et~al.} 2021,
  Philosophical Transactions of the Royal Society of London Series A, 379,
  20200171, \dodoi{10.1098/rsta.2020.0171}

\bibitem[{{Maltby} {et~al.}(1986){Maltby}, {Avrett}, {Carlsson},
  {Kjeldseth-Moe}, {Kurucz}, \& {Loeser}}]{1986ApJ...306..284M}
{Maltby}, P., {Avrett}, E.~H., {Carlsson}, M., {et~al.} 1986, \apj, 306, 284,
  \dodoi{10.1086/164342}

\bibitem[{{Marino} {et~al.}(2006){Marino}, {Rimmele}, \&
  {Christou}}]{2006SPIE.6272E..3WM}
{Marino}, J., {Rimmele}, T., \& {Christou}, J. 2006, in Society of
  Photo-Optical Instrumentation Engineers (SPIE) Conference Series, Vol. 6272,
  Society of Photo-Optical Instrumentation Engineers (SPIE) Conference Series,
  ed. B.~L. {Ellerbroek} \& D.~{Bonaccini Calia}, 62723W,
  \dodoi{10.1117/12.672297}

\bibitem[{{Monson} {et~al.}(2021){Monson}, {Mathioudakis}, {Reid}, {Milligan},
  \& {Kuridze}}]{Monson2021}
{Monson}, A.~J., {Mathioudakis}, M., {Reid}, A., {Milligan}, R., \& {Kuridze},
  D. 2021, \apj, 915, 16, \dodoi{10.3847/1538-4357/abfda8}

\bibitem[{{Moreels} {et~al.}(2013){Moreels}, {Goossens}, \& {Van
  Doorsselaere}}]{Moreels2013b}
{Moreels}, M.~G., {Goossens}, M., \& {Van Doorsselaere}, T. 2013, \aap, 555,
  A75, \dodoi{10.1051/0004-6361/201321545}

\bibitem[{{Moreels} \& {Van Doorsselaere}(2013)}]{Moreels2013a}
{Moreels}, M.~G., \& {Van Doorsselaere}, T. 2013, \aap, 551, A137,
  \dodoi{10.1051/0004-6361/201219568}

\bibitem[{{Moreels} {et~al.}(2015){Moreels}, {Van Doorsselaere}, {Grant},
  {Jess}, \& {Goossens}}]{Moreels2015}
{Moreels}, M.~G., {Van Doorsselaere}, T., {Grant}, S.~D.~T., {Jess}, D.~B., \&
  {Goossens}, M. 2015, \aap, 578, A60, \dodoi{10.1051/0004-6361/201425468}

\bibitem[{{Morton} {et~al.}(2011){Morton}, {Erd{\'e}lyi}, {Jess}, \&
  {Mathioudakis}}]{Morton2011}
{Morton}, R.~J., {Erd{\'e}lyi}, R., {Jess}, D.~B., \& {Mathioudakis}, M. 2011,
  \apjl, 729, L18, \dodoi{10.1088/2041-8205/729/2/L18}

\bibitem[{{Narain} \& {Ulmschneider}(1996)}]{Narain1996}
{Narain}, U., \& {Ulmschneider}, P. 1996, \ssr, 75, 453,
  \dodoi{10.1007/BF00833341}

\bibitem[{{Nelson} {et~al.}(2017){Nelson}, {Henriques}, {Mathioudakis}, \&
  {Keenan}}]{Nelson2017}
{Nelson}, C.~J., {Henriques}, V.~M.~J., {Mathioudakis}, M., \& {Keenan}, F.~P.
  2017, \aap, 605, A14, \dodoi{10.1051/0004-6361/201730467}

\bibitem[{{Nordlund} \& {Stein}(2001)}]{Nord2001}
{Nordlund}, {\r{A}}., \& {Stein}, R.~F. 2001, \apj, 546, 576,
  \dodoi{10.1086/318217}

\bibitem[{{Nordlund} {et~al.}(2009){Nordlund}, {Stein}, \&
  {Asplund}}]{Nord2009}
{Nordlund}, {\r{A}}., {Stein}, R.~F., \& {Asplund}, M. 2009, Living Reviews in
  Solar Physics, 6, 2, \dodoi{10.12942/lrsp-2009-2}

\bibitem[{{Noyes} \& {Leighton}(1963)}]{Noyes1963}
{Noyes}, R.~W., \& {Leighton}, R.~B. 1963, \apj, 138, 631,
  \dodoi{10.1086/147675}

\bibitem[{{Parnell} \& {De Moortel}(2012)}]{Parnell2012}
{Parnell}, C.~E., \& {De Moortel}, I. 2012, Philosophical Transactions of the
  Royal Society of London Series A, 370, 3217, \dodoi{10.1098/rsta.2012.0113}

\bibitem[{{Priest} {et~al.}(2018){Priest}, {Chitta}, \&
  {Syntelis}}]{2018ApJ...862L..24P}
{Priest}, E.~R., {Chitta}, L.~P., \& {Syntelis}, P. 2018, \apjl, 862, L24,
  \dodoi{10.3847/2041-8213/aad4fc}

\bibitem[{{Rajaguru} {et~al.}(2019){Rajaguru}, {Sangeetha}, \&
  {Tripathi}}]{Rajaguru2019}
{Rajaguru}, S.~P., {Sangeetha}, C.~R., \& {Tripathi}, D. 2019, \apj, 871, 155,
  \dodoi{10.3847/1538-4357/aaf883}

\bibitem[{{Rast} {et~al.}(2021){Rast}, {Bello Gonz{\'a}lez}, {Bellot Rubio},
  {Cao}, {Cauzzi}, {Deluca}, {de Pontieu}, {Fletcher}, {Gibson}, {Judge},
  {Katsukawa}, {Kazachenko}, {Khomenko}, {Landi}, {Mart{\'\i}nez Pillet},
  {Petrie}, {Qiu}, {Rachmeler}, {Rempel}, {Schmidt}, {Scullion}, {Sun},
  {Welsch}, {Andretta}, {Antolin}, {Ayres}, {Balasubramaniam}, {Ballai},
  {Berger}, {Bradshaw}, {Campbell}, {Carlsson}, {Casini}, {Centeno}, {Cranmer},
  {Criscuoli}, {Deforest}, {Deng}, {Erd{\'e}lyi}, {Fedun}, {Fischer},
  {Gonz{\'a}lez Manrique}, {Hahn}, {Harra}, {Henriques}, {Hurlburt}, {Jaeggli},
  {Jafarzadeh}, {Jain}, {Jefferies}, {Keys}, {Kowalski}, {Kuckein}, {Kuhn},
  {Kuridze}, {Liu}, {Liu}, {Longcope}, {Mathioudakis}, {McAteer}, {McIntosh},
  {McKenzie}, {Miralles}, {Morton}, {Muglach}, {Nelson}, {Panesar}, {Parenti},
  {Parnell}, {Poduval}, {Reardon}, {Reep}, {Schad}, {Schmit}, {Sharma},
  {Socas-Navarro}, {Srivastava}, {Sterling}, {Suematsu}, {Tarr}, {Tiwari},
  {Tritschler}, {Verth}, {Vourlidas}, {Wang}, {Wang}, {NSO and DKIST Project},
  {DKIST Instrument Scientists}, {DKIST Science Working Group}, \& {DKIST
  Critical Science Plan Community}}]{2021SoPh..296...70R}
{Rast}, M.~P., {Bello Gonz{\'a}lez}, N., {Bellot Rubio}, L., {et~al.} 2021,
  \solphys, 296, 70, \dodoi{10.1007/s11207-021-01789-2}

\bibitem[{{Riedl} {et~al.}(2021){Riedl}, {Gilchrist-Millar}, {Van
  Doorsselaere}, {Jess}, \& {Grant}}]{Riedl2021}
{Riedl}, J.~M., {Gilchrist-Millar}, C.~A., {Van Doorsselaere}, T., {Jess},
  D.~B., \& {Grant}, S.~D.~T. 2021, \aap, 648, A77,
  \dodoi{10.1051/0004-6361/202040163}

\bibitem[{{Rimmele}(2004)}]{Rimmele2004}
{Rimmele}, T.~R. 2004, in Society of Photo-Optical Instrumentation Engineers
  (SPIE) Conference Series, Vol. 5490, Advancements in Adaptive Optics, ed.
  D.~{Bonaccini Calia}, B.~L. {Ellerbroek}, \& R.~{Ragazzoni}, 34--46,
  \dodoi{10.1117/12.551764}

\bibitem[{{Rimmele} {et~al.}(2020){Rimmele}, {Warner}, {Keil}, {Goode},
  {Kn{\"o}lker}, {Kuhn}, {Rosner}, {McMullin}, {Casini}, {Lin}, {W{\"o}ger},
  {von der L{\"u}he}, {Tritschler}, {Davey}, {de Wijn}, {Elmore}, {Fehlmann},
  {Harrington}, {Jaeggli}, {Rast}, {Schad}, {Schmidt}, {Mathioudakis},
  {Mickey}, {Anan}, {Beck}, {Marshall}, {Jeffers}, {Oschmann}, {Beard},
  {Berst}, {Cowan}, {Craig}, {Cross}, {Cummings}, {Donnelly}, {de Vanssay},
  {Eigenbrot}, {Ferayorni}, {Foster}, {Galapon}, {Gedrites}, {Gonzales},
  {Goodrich}, {Gregory}, {Guzman}, {Guzzo}, {Hegwer}, {Hubbard}, {Hubbard},
  {Johansson}, {Johnson}, {Liang}, {Liang}, {McQuillen}, {Mayer}, {Newman},
  {Onodera}, {Phelps}, {Puentes}, {Richards}, {Rimmele}, {Sekulic}, {Shimko},
  {Simison}, {Smith}, {Starman}, {Sueoka}, {Summers}, {Szabo}, {Szabo},
  {Wampler}, {Williams}, \& {White}}]{Rim2020}
{Rimmele}, T.~R., {Warner}, M., {Keil}, S.~L., {et~al.} 2020, \solphys, 295,
  172, \dodoi{10.1007/s11207-020-01736-7}

\bibitem[{{Rouppe van der Voort} \& {de la Cruz
  Rodr{\'\i}guez}(2013)}]{Rouppe2013}
{Rouppe van der Voort}, L., \& {de la Cruz Rodr{\'\i}guez}, J. 2013, \apj, 776,
  56, \dodoi{10.1088/0004-637X/776/1/56}

\bibitem[{{Ruiz Cobo} \& {del Toro Iniesta}(1992)}]{Ruiz1992}
{Ruiz Cobo}, B., \& {del Toro Iniesta}, J.~C. 1992, \apj, 398, 375,
  \dodoi{10.1086/171862}

\bibitem[{{Schmitz} \& {Fleck}(1998)}]{Schmitz1998}
{Schmitz}, F., \& {Fleck}, B. 1998, \aap, 337, 487

\bibitem[{{Schwarzschild}(1948)}]{Sch1948}
{Schwarzschild}, M. 1948, \apj, 107, 1, \dodoi{10.1086/144983}

\bibitem[{{Severino} {et~al.}(2013){Severino}, {Straus}, {Oliviero}, {Steffen},
  \& {Fleck}}]{Severino2013}
{Severino}, G., {Straus}, T., {Oliviero}, M., {Steffen}, M., \& {Fleck}, B.
  2013, \solphys, 284, 297, \dodoi{10.1007/s11207-012-0172-9}

\bibitem[{{Snow} {et~al.}(2018){Snow}, {Fedun}, {Gent}, {Verth}, \&
  {Erd{\'e}lyi}}]{Snow2018}
{Snow}, B., {Fedun}, V., {Gent}, F.~A., {Verth}, G., \& {Erd{\'e}lyi}, R. 2018,
  \apj, 857, 125, \dodoi{10.3847/1538-4357/aab7f7}

\bibitem[{{Sobotka}(2003)}]{Sob2003}
{Sobotka}, M. 2003, Astronomische Nachrichten, 324, 369,
  \dodoi{10.1002/asna.200310132}

\bibitem[{{Socas-Navarro}(2005)}]{Socas2005}
{Socas-Navarro}, H. 2005, \apjl, 633, L57, \dodoi{10.1086/498145}

\bibitem[{{Socas-Navarro} {et~al.}(2000){Socas-Navarro}, {Trujillo Bueno}, \&
  {Ruiz Cobo}}]{Socas2000}
{Socas-Navarro}, H., {Trujillo Bueno}, J., \& {Ruiz Cobo}, B. 2000, Science,
  288, 1396, \dodoi{10.1126/science.288.5470.1396}

\bibitem[{{Solanki} {et~al.}(2017){Solanki}, {Riethm{\"u}ller}, {Barthol},
  {Danilovic}, {Deutsch}, {Doerr}, {Feller}, {Gandorfer}, {Germerott}, {Gizon},
  {Grauf}, {Heerlein}, {Hirzberger}, {Kolleck}, {Lagg}, {Meller}, {Tomasch},
  {van Noort}, {Blanco Rodr{\'\i}guez}, {Gasent Blesa}, {Balaguer Jim{\'e}nez},
  {Del Toro Iniesta}, {L{\'o}pez Jim{\'e}nez}, {Orozco Suarez}, {Berkefeld},
  {Halbgewachs}, {Schmidt}, {{\'A}lvarez-Herrero}, {Sabau-Graziati}, {P{\'e}rez
  Grande}, {Mart{\'\i}nez Pillet}, {Card}, {Centeno}, {Kn{\"o}lker}, \&
  {Lecinski}}]{Solanki2017}
{Solanki}, S.~K., {Riethm{\"u}ller}, T.~L., {Barthol}, P., {et~al.} 2017,
  \apjs, 229, 2, \dodoi{10.3847/1538-4365/229/1/2}

\bibitem[{{Srivastava} {et~al.}(2021){Srivastava}, {Ballester}, {Cally},
  {Carlsson}, {Goossens}, {Jess}, {Khomenko}, {Mathioudakis}, {Murawski}, \&
  {Zaqarashvili}}]{2021JGRA..12629097S}
{Srivastava}, A.~K., {Ballester}, J.~L., {Cally}, P.~S., {et~al.} 2021, Journal
  of Geophysical Research (Space Physics), 126, e29097,
  \dodoi{10.1029/2020JA029097}

\bibitem[{{Stangalini} {et~al.}(2011){Stangalini}, {Del Moro}, {Berrilli}, \&
  {Jefferies}}]{Stan2011}
{Stangalini}, M., {Del Moro}, D., {Berrilli}, F., \& {Jefferies}, S.~M. 2011,
  \aap, 534, A65, \dodoi{10.1051/0004-6361/201117356}

\bibitem[{{Stangalini} {et~al.}(2012){Stangalini}, {Giannattasio}, {Del Moro},
  \& {Berrilli}}]{Stan2012}
{Stangalini}, M., {Giannattasio}, F., {Del Moro}, D., \& {Berrilli}, F. 2012,
  \aap, 539, L4, \dodoi{10.1051/0004-6361/201118654}

\bibitem[{{Stangalini} {et~al.}(2021){Stangalini}, {Jess}, {Verth}, {Fedun},
  {Fleck}, {Jafarzadeh}, {Keys}, {Murabito}, {Calchetti}, {Aldhafeeri},
  {Berrilli}, {Del Moro}, {Jefferies}, {Terradas}, \&
  {Soler}}]{2021A&A...649A.169S}
{Stangalini}, M., {Jess}, D.~B., {Verth}, G., {et~al.} 2021, \aap, 649, A169,
  \dodoi{10.1051/0004-6361/202140429}

\bibitem[{{Stangalini} {et~al.}(2022){Stangalini}, {Verth}, {Fedun},
  {Aldhafeeri}, {Jess}, {Jafarzadeh}, {Keys}, {Fleck}, {Terradas}, {Murabito},
  {Ermolli}, {Soler}, {Giorgi}, \& {Macbride}}]{Stan2022}
{Stangalini}, M., {Verth}, G., {Fedun}, V., {et~al.} 2022, Nature
  Communications, 13, 479, \dodoi{10.1038/s41467-022-28136-8}

\bibitem[{{Stein} \& {Nordlund}(2001)}]{Stein2001}
{Stein}, R.~F., \& {Nordlund}, {\r{A}}. 2001, \apj, 546, 585,
  \dodoi{10.1086/318218}

\bibitem[{{Steiner} {et~al.}(1998){Steiner}, {Grossmann-Doerth}, {Kn{\"o}lker},
  \& {Sch{\"u}ssler}}]{Steiner1998}
{Steiner}, O., {Grossmann-Doerth}, U., {Kn{\"o}lker}, M., \& {Sch{\"u}ssler},
  M. 1998, \apj, 495, 468, \dodoi{10.1086/305255}

\bibitem[{{Stull}(1988)}]{StullRolandB1988AItB}
{Stull}, R.~B. 1988, Atmospheric and Oceanographic Sciences Library, Vol.~13,
  An Introduction to Boundary Layer Meteorology (Dordrecht: Springer
  Netherlands)

\bibitem[{{Torrence} \& {Compo}(1998)}]{Torrence1998}
{Torrence}, C., \& {Compo}, G.~P. 1998, Bulletin of the American Meteorological
  Society, 79, 61, \dodoi{10.1175/1520-0477(1998)079<0061:APGTWA>2.0.CO;2}

\bibitem[{{Torrence} \& {Webster}(1999)}]{Torrence1999}
{Torrence}, C., \& {Webster}, P.~J. 1999, Journal of Climate, 12, 2679,
  \dodoi{10.1175/1520-0442(1999)012<2679:ICITEM>2.0.CO;2}

\bibitem[{{Tsap} {et~al.}(2016){Tsap}, {Stepanov}, \& {Kopylova}}]{Tsap2016}
{Tsap}, Y.~T., {Stepanov}, A.~V., \& {Kopylova}, Y.~G. 2016, \solphys, 291,
  3349, \dodoi{10.1007/s11207-016-0980-4}

\bibitem[{{Uitenbroek}(2006)}]{Uit2006}
{Uitenbroek}, H. 2006, \apj, 639, 516, \dodoi{10.1086/499220}

\bibitem[{{Van Doorsselaere} {et~al.}(2020){Van Doorsselaere}, {Srivastava},
  {Antolin}, {Magyar}, {Vasheghani Farahani}, {Tian}, {Kolotkov}, {Ofman},
  {Guo}, {Arregui}, {De Moortel}, \& {Pascoe}}]{VD2020}
{Van Doorsselaere}, T., {Srivastava}, A.~K., {Antolin}, P., {et~al.} 2020,
  \ssr, 216, 140, \dodoi{10.1007/s11214-020-00770-y}

\bibitem[{{Vernazza} {et~al.}(1981){Vernazza}, {Avrett}, \&
  {Loeser}}]{Vern1981}
{Vernazza}, J.~E., {Avrett}, E.~H., \& {Loeser}, R. 1981, \apjs, 45, 635,
  \dodoi{10.1086/190731}

\bibitem[{{Verth} \& {Jess}(2016)}]{2016GMS...216..431V}
{Verth}, G., \& {Jess}, D.~B. 2016, Washington DC American Geophysical Union
  Geophysical Monograph Series, 216, 431, \dodoi{10.1002/9781119055006.ch25}

\bibitem[{{Wilson} \& {Cannon}(1968)}]{Wilson1968}
{Wilson}, P.~R., \& {Cannon}, C.~J. 1968, \solphys, 4, 3,
  \dodoi{10.1007/BF00146994}

\bibitem[{{Wi{\'s}niewska} {et~al.}(2016){Wi{\'s}niewska}, {Musielak},
  {Staiger}, \& {Roth}}]{Wis2016}
{Wi{\'s}niewska}, A., {Musielak}, Z.~E., {Staiger}, J., \& {Roth}, M. 2016,
  \apjl, 819, L23, \dodoi{10.3847/2041-8205/819/2/L23}

\bibitem[{{Withbroe} \& {Noyes}(1977)}]{Withbroe1977}
{Withbroe}, G.~L., \& {Noyes}, R.~W. 1977, \araa, 15, 363,
  \dodoi{10.1146/annurev.aa.15.090177.002051}

\bibitem[{{W{\"o}ger} {et~al.}(2008){W{\"o}ger}, {von der L{\"u}he}, \&
  {Reardon}}]{Woger2008}
{W{\"o}ger}, F., {von der L{\"u}he}, O., \& {Reardon}, K. 2008, \aap, 488, 375,
  \dodoi{10.1051/0004-6361:200809894}

\bibitem[{{Yuan} {et~al.}(2014){Yuan}, {Nakariakov}, {Huang}, {Li}, {Su},
  {Yan}, \& {Tan}}]{2014ApJ...792...41Y}
{Yuan}, D., {Nakariakov}, V.~M., {Huang}, Z., {et~al.} 2014, \apj, 792, 41,
  \dodoi{10.1088/0004-637X/792/1/41}

\bibitem[{{Yurchyshyn} {et~al.}(2014){Yurchyshyn}, {Abramenko}, {Kosovichev},
  \& {Goode}}]{Yurch2014}
{Yurchyshyn}, V., {Abramenko}, V., {Kosovichev}, A., \& {Goode}, P. 2014, \apj,
  787, 58, \dodoi{10.1088/0004-637X/787/1/58}

\bibitem[{{Zwaan}(1985)}]{Zwaan1985}
{Zwaan}, C. 1985, \solphys, 100, 397, \dodoi{10.1007/BF00158438}

\end{thebibliography}

\end{document}